\documentclass[12pt, letterpaper]{article}

\usepackage{amsmath}
\usepackage{amsfonts}
\usepackage{amssymb}
\usepackage{graphicx}

\usepackage{color}

\usepackage{amsmath, amssymb, graphics, epsfig, graphicx}
\usepackage{epsf}
\usepackage{epstopdf}
\usepackage {amssymb}
\newcommand{\nc}{\newcommand}
\nc{\ba}{\begin{eqnarray}}
\nc{\ea}{\end{eqnarray}}
\newcommand\be{\begin{equation}}
\newcommand\ee{\end{equation}}

\newcommand{\calR}{{\cal{R}}}

\newcommand{\bfx}{{\bf{x}}}

\begin{document}

%%%%%%%%%%%%%%%%%%%%%%%%%%%%%%%%%%%%%%%%%%%%%%%%%%%%%
%\begin{flushright} {\footnotesize IC/2007/001}  \end{flushright}
\vspace{5mm}
\vspace{0.5cm}
\begin{center}

\def\thefootnote{\fnsymbol{footnote}}

{\Large  Higher Derivative Mimetic Gravity}
\\[0.5cm]

{  Mohammad Ali Gorji$^{1}\footnote{gorji@ipm.ir}$, Seyed  Ali Hosseini 
	Mansoori$^{ 2, 1}\footnote{ shosseini@shahroodut.ac.ir, shossein@ipm.ir}$,  
	Hassan Firouzjahi$^{1}\footnote{firouz@ipm.ir }$}
\\[0.5cm]

{\small \textit{$^1$School of Astronomy, Institute for Research in Fundamental Sciences (IPM) \\ P.~O.~Box 19395-5531, Tehran, Iran
}}\\

{\small \textit{$^2$
		Physics Department, Shahrood University of Technology,\\ P.O.Box 3619995161 Shahrood, Iran }}
\end{center}

\vspace{.8cm}

\hrule \vspace{0.3cm}
%{\small  \noindent \textbf{Abstract} \\[0.3cm]
%\noindent

%%%%%%%%%%%%%%%%%%%%%%%%%%%%%%%%%%%%%%%%%%%%%%%%%%%%%

\begin{abstract}

We study cosmological perturbations in mimetic gravity in the presence of classified higher 
derivative terms which can make the mimetic perturbations stable. We show that the quadratic higher 
derivative terms which are independent of curvature and the cubic higher derivative terms which come 
from curvature corrections are sufficient to remove  instabilities in mimetic perturbations. The classified 
higher derivative terms have the same dimensions but they contribute differently in the background and 
perturbed equations. Therefore,  we can control both the background and the perturbation equations 
allowing us to construct the  higher derivative extension of mimetic dark matter and the mimetic 
nonsingular bouncing scenarios.  The latter can be thought as a new higher derivative effective action 
for the loop quantum cosmology  scenario in which the equations of motion coincide with those 
suggested by loop quantum cosmology.  We investigate a possible connection between the mimetic 
cosmology and the Randall-Sundrum cosmology. 

\end{abstract}
\vspace{0.5cm} \hrule
\def\thefootnote{\arabic{footnote}}
\setcounter{footnote}{0}
\newpage
%%%%%%%%%%%%%%%%%%%%%%%%%%%%%%%%%%%%%%%%%%%%%%%%%%%%%
\section{Introduction}

General relativity has been experimentally confirmed on distances at the order of the solar system, but  
it may be modified on cosmological scales. For instance, the late time comic acceleration and the 
observed galaxy rotation curves may not be explained by general relativity coupled to the known matters 
which lead to the so-called dark energy and dark matter problems. These problems may be addressed 
partially in the context of scalar-tensor theories in which one couples a scalar field to gravity. The 
nature of the scalar field in these models is however a matter of question. 

Recently a new scalar-tensor model, known as 
the mimetic gravity, has been suggested in
\cite{Chamseddine:2013kea} in which the scalar field  plays the role of dark matter. The idea is to consider 
a metric transformation\footnote{We use the metric signature convention $(-, +, +, +)$.} $g_{\mu \nu} = 
- (\tilde g^{\alpha\beta}\partial_\alpha \phi \partial_\beta \phi ) \tilde g_{\mu \nu}$ in which $g_{\mu\nu}$ is 
the ``physical'' metric which is related to an auxiliary metric $\tilde g_{\mu \nu}$ via the scalar field $\phi$. 
This transformation separates the conformal degree of freedom of gravitational field in a covariant manner 
\cite{Bekenstein:1992pj}. The new degree of freedom $\phi$ makes the longitudinal mode of gravity 
dynamical even in the absence of conventional standard matter fields \cite{Chamseddine:2013kea}. The 
above transformation in mimetic scenario can also be seen as a particular limit of disformal 
transformations \cite{Deruelle:2014zza}. More precisely, the number of degrees of freedom remains 
unchanged under a general invertible disformal transformation \cite{Domenech:2015tca, Arroja:2015wpa, Motohashi:2015pra, Takahashi:2017zgr}. 
But, the mimetic scenario can be realized from the singular (non-invertible) limit of the disformal 
transformation and therefore a new degree of freedom $\phi$ arises in this setup 
\cite{Arroja:2015wpa,Domenech:2015tca,Achour:2016rkg}. Moreover, it is shown that the mimetic scenario 
can also be interpreted as an infrared limit of Horava-Lifshitz gravity \cite{Ramazanov:2016xhp}.  
For recent developments and other studies in mimetic gravity see \cite{MimeticNew}.

The action of mimetic gravity can be  written in term of physical metric $g_{\mu \nu}$ and the scalar field 
as\footnote{We work in natural unit in which $M_{_{\rm Pl}}=1$ where $M_{_{\rm Pl}}$ is the reduced 
Planck mass.} \cite{Golovnev:2013jxa}
\ba\label{action0}
S= \int d^4 x \sqrt{-g} \left[\frac{1}{2} R + \lambda (X+1) \right] \,,
\ea
in which $X$ is defined as
\ba\label{X}
X=g^{\mu\nu}\partial_{\mu}\phi \partial_{\nu}\phi \, ,
\ea
and the auxiliary field $\lambda$ enforces the mimetic constraint
\ba\label{mimetic-C}
X = -1 \, . 
\ea

The new degree of freedom $\phi$ provides the scalar mode which is present even in the 
absence of ordinary matter source. Interestingly, it mimics the roles of dark matter in 
cosmological background and this is the reason for which the scenario is called mimetic dark 
matter. At the level of perturbations, the model is free from instabilities \cite{Barvinsky:2013mea} 
but the sound speed of perturbations turns out to be zero. Therefore, a higher derivative 
term proportional to $(\Box\phi)^2$ is added to the action (\ref{action0}) which generates  
a nonzero and constant sound speed for the perturbations \cite{Chamseddine:2014vna}. 
Even in the presence of the higher derivative term  $(\Box\phi)^2$ the mimetic model still 
describes a dark matter dominated universe \cite{Mirzagholi:2014ifa}, but this model 
suffers from the ghost and the gradient instabilities \cite{Ramazanov:2016xhp,Ijjas:2016pad}. 

On the other hand, in recent papers \cite{Chamseddine:2016ktu,Chamseddine:2016uef}, the  
setup (\ref{action0}) is modified such that the quadratic higher derivative term $(\Box\phi)^2$ 
was replaced with a general higher derivative function $f(\Box\phi)$ as follows 
\ba\label{action1}
S= \int d^4 x  \sqrt{-g} \left[ \frac{1}{2} R + \lambda \left(X+ 1 \right) 
+ f(\Box\phi)\right] \, .
\ea
 In an FRW background 
\ba\label{FRW}
ds^2 = - dt^2 + a(t)^2 d \bfx^2 \, ,
\ea
the mimetic constraint equation (\ref{mimetic-C}) gives 
\ba\label{phi}
\dot \phi=1 \, , \hspace{1cm} \phi = t \, ,
\ea
and the background field equations from the action (\ref{action1}) is obtained to be
\ba
\label{Friedmann-GHD}
3 H^2=-2\bar{\lambda} -(f+3Hf_{\chi}+3\dot{H}f_{\chi\chi})\,,
\ea
and
\ba
\label{Raychuadhuri-GHD}
2\dot{H}+3H^2=-(f+3Hf_{\chi}-3\dot{H}f_{\chi\chi})\, ,
\ea
in which $H= \dot a(t)/a(t)$ is the Hubble expansion rate, 
$\bar{\lambda}$ denotes the background value of $\lambda$ and $f_{\chi}$ denotes a 
derivative with respect to $\chi \equiv \Box \phi$ and so on.

It was shown in \cite{Chamseddine:2016ktu,Chamseddine:2016uef} that the curvature takes a 
maximum value for the special form of  function $f(\chi)$  (see Eq. (\ref{f}) below) and the 
cosmological big bang singularity and the singularity of Schwarzschild black hole can be resolved in 
this framework. Also, the background cosmological equations in this model coincide with the 
bouncing equations suggested by loop quantum cosmology \cite{Bodendorfer:2017bjt,Liu:2017puc}. 
This coincidence is important since the modified Friedmann equations  in loop quantum cosmology 
scenario are not deduced from a covariant action but they are obtained at the semiclassical regime 
of the quantized loop scenario \cite{LQC1,LQC2,LQC3}. However, it is shown that these equations 
can be achieved by adding some higher derivative terms to the action of standard classical general 
relativity \cite{Mukhanov:1991zn,Brandenberger:1993ef,Date:2008gq,Qiu:2011cy,Yoshida:2017swb}. 
In this  respect the mimetic scenario  suggests a higher derivative scalar tensor model for the 
effective Friedmann equations of loop quantum cosmology \cite{Bodendorfer:2017bjt,Liu:2017puc}. 

 The action (\ref{action1}) thus describes both the mimetic dark matter 
\cite{Chamseddine:2013kea,Chamseddine:2014vna} and mimetic nonsingular model 
\cite{Chamseddine:2016uef,Chamseddine:2016ktu} for different functional forms of 
$f(\Box{\phi})$.  Apart from these interesting features, the scenario still suffers from ghost 
and gradient instabilities \cite{Firouzjahi:2017txv}. However, using the effective field theory 
methods, it is found that considering a coupling between the second derivative of $\phi$ 
and the Ricci scalar and Ricci tensor would make the mimetic setup stable and a healthy 
extension of mimetic dark matter is then suggested in \cite{Hirano:2017zox} (see also Refs. 
\cite{Cai:2017dyi,Cai:2017dxl} in which bouncing solutions in the context of effective field 
theory are studied). On the other hand, covariant stable extension of the mimetic model 
(\ref{action0}) is suggested in which the higher derivative terms $f(\Box\phi) R$, 
$(\Box\phi)^2$ and $\nabla_\mu\nabla_\nu\phi\nabla^\mu\nabla^\nu\phi$ are considered 
\cite{Zheng:2017qfs}. Although adding these terms to the action (\ref{action0}) can make 
the setup stable, but the background equations no longer describe a dark matter-like fluid. 
In the same manner, one can add the curvature dependent higher derivative terms like 
$f(\Box\phi) R$ to the action of nonsingular mimetic  bouncing model (\ref{action1}) in order 
to remove the instabilities. But, again the background equations will differ from 
(\ref{Friedmann-GHD}) and (\ref{Raychuadhuri-GHD}) and it is not clear how the singularity 
can be removed in the resultant model. In this paper, we work with the classified higher 
derivative terms like $(\Box\phi)^2$ and $\nabla_\mu\nabla_\nu\phi\nabla^\mu\nabla^\nu\phi$ 
(or ${\Box\phi }R$ and $\nabla_\mu\nabla_\nu\phi R_{\mu\nu}$) which have the same 
dimensions. The advantage of working with these classified terms is that they 
appear in different combinations in the background and perturbed equations. This allow us to 
control both the background and the perturbed equations by appropriately choosing the 
coefficients of the higher derivative terms. In comparison with Ref. \cite{Hirano:2017zox}
where the authors find healthy mimetic dark matter through the effective field theory 
methods, we can find covariant action for the mimetic dark matter and also mimetic 
bouncing scenario. Moreover, we allow all the possible higher derivative terms 
including Riemann tensor,  in addition to Ricci scalar and Ricci tensor which were considered 
in Ref. \cite{Hirano:2017zox} in the context of effective field theory. 

The structure of the paper is as follows. In the next section, we introduce a simple healthy 
mimetic model by means of curvature independent quadratic  higher derivative terms and 
curvature dependent cubic higher derivative terms. In section 3, we introduce new mimetic 
model by adding quintic curvature dependent higher derivative terms to the action. Using 
this model, in section 4 we construct new mimetic dark matter and mimetic bouncing 
scenarios. We also comment on the possible connection between mimetic scenario and 
Randall-Sundrum cosmology. The last section is devoted to summaries and conclusions. 
Some messy formula and details of calculations are shown in appendix A. In appendix B, 
we explain why in section 3 we have not considered the simple cases of cubic and quartic 
curvature dependent terms instead of the quintic terms.

%%%%%%%%%%%%%%%%%%%%%%%%%%%%%%%%%%%%%%%%%%%%%%%%%%%%%
\section{The stable higher order mimetic gravity}\label{setup-sec}

In this section, we introduce our method in which not only the instabilities in mimetic perturbations are 
removed but also we can have a  control on the background equations. 

In order to control the background equations, we employ some classified higher derivative terms. More 
precisely, we should first find higher derivative terms with the same dimensions and then add them with 
different coefficients to the action. They contributes  differently  in the  background and perturbed 
equations. This allows us to  remove the instabilities in mimetic perturbations and at the same time 
to  keep the background equations as  in Eqs. (\ref{Friedmann-GHD}) and (\ref{Raychuadhuri-GHD}).

The mimetic scenario is a scalar-tensor theory and, fortunately, the classification of higher 
derivative terms was previously done in the context of higher derivative scalar-tensor theories 
such as Horndeski models \cite{Horndeski:1974wa,Nicolis:2008in,Deffayet:2009wt}. Indeed, in 
the absence of any primary constraint, the higher derivative terms such as $f(\Box\phi)$ would 
lead to the Ostrogradsky ghost \cite{Ostrogradsky:1850fid,Woodard:2015zca} and therefore in 
Horndeski theories one fixes the coefficients of the classified higher derivative terms such that 
the theory being free of higher derivative ghosts. Indeed, in Horndeski theories, the equation of 
motions are second order but the action contains higher derivative terms. Thus, one can work 
with the on-shell action in which the equation of motion is imposed and therefore it does not 
contain higher derivative terms. However, this is  not the only possibility to avoid Ostrogradsky
ghost. Recently, the theories known as beyond Horndeski have been suggested in which the
equation of motions are no longer second order but the theories are still free of Ostrogradsky 
ghost \cite{Gleyzes:2014dya}. Fixing the coefficients in theories such as Horndeski and beyond 
Horndeski is equivalent to imposing some second class constraints to the system under 
consideration. This in turn reduces the number of degrees of freedom and therefore prevents 
the propagation of Ostrogradsky-like ghosts \cite{Langlois:2015cwa, Crisostomi:2016czh, Crisostomi:2016tcp}.

 In comparison with the Horndeski and beyond Horndeski models, there is an additional 
primary constraint in mimetic scenario {\it i.e.} the mimetic constraint (\ref{mimetic-C}) which 
allows us to consider the cubic higher derivative terms like $\Box\phi R$ and 
$\phi^{\mu\nu}R_{\mu\nu}$ with arbitrary coefficients as we have considered in (\ref{action}). 
Note that these terms only appear with the fixed combination $\left(R_{\mu\nu}-\frac{1}{2}
g_{\mu\nu} R\right) \phi^{\mu\nu}$ in Horndeski models. In the same manner, the arbitrary 
combination of the quadratic terms $(\Box\phi)^2$ and $\nabla_\mu\nabla_\nu\phi
\nabla^\mu\nabla^\nu\phi$ would generally produce Ostrogradsky ghost and therefore they 
always appear with the fixed combination $(\Box\phi)^2
-\nabla_\mu\nabla_\nu\phi\nabla^\mu\nabla^\nu\phi$  in Horndeski scenario 
\cite{Arroja:2015wpa,Arroja:2015yvd,Cognola:2016gjy,Arroja:2017msd,Takahashi:2017pje}, while 
one can consider these terms with general coefficients in mimetic scenario \cite{Zheng:2017qfs}. 
Indeed, the mimetic constraint (\ref{mimetic-C}) separately makes these terms free of 
Ostrogradsky ghosts. Therefore, in higher order mimetic gravity model (\ref{action}), we have 
implemented these classified higher derivative terms with general coefficients. We then fix the 
coefficients in such a way that the background equations remain unchanged, i.e. they keep the 
forms of (\ref{Friedmann-GHD}) and (\ref{Raychuadhuri-GHD}). In the next step, we further 
constrain the coefficients such that the setup be stable under cosmological perturbations.

%%%%%%%%%%%%%%%%%%%%%%%%%%%%%%%%%%%%%%%%%%%%%%%%%%%%%%%%%
\subsection{The model}

Following the above discussions, we consider the higher derivative extension of the mimetic model with 
the action 
\ba
\label{action}
S= \int d^4 x  \sqrt{-g} \left[ \frac{1}{2} R + \lambda \left(X+ 1 \right) - V(\phi)
+ \sum_{i=1}^{5} \Big(\alpha_i L_i^{(2,0)} + \kappa_i L_i^{(1,2)}\Big) \right] \, ,
\ea
in which $\alpha_i$ are the coefficients of the five quadratic higher derivative terms $L_i^{(2,0)}$ which 
are independent of curvature,  defined as \cite{Achour:2016rkg}
\ba\label{L2i}
&&L_1^{(2,0)} = \phi_{\mu\nu} \phi^{\mu\nu} \, , \hspace{1cm} L_2^{(2,0)} = (\Box\phi)^2 \, , 
\hspace{1cm} L_3^{(2,0)} = (\Box\phi) \phi^\mu \phi_{\mu\nu} \phi^{\nu} \, , \nonumber \\  
&&L_4^{(2,0)} = \phi^{\mu} \phi_{\mu\rho} \phi^{\rho\nu} \phi_{\nu} \, , \hspace{1cm}
L_5^{(2,0)} =  (\phi^\mu \phi_{\mu\nu} \phi^{\nu})^2 \, ,
\ea
and $\kappa_i$ are the coefficients of the five cubic higher derivative terms $L_i^{(1,2)}$, constructed 
from curvature, given by (see appendix A of Ref. \cite{BenAchour:2016fzp} for the details)
\ba\label{L3i}
&&L_1^{(1,2)} =  \phi^{\mu\nu} \, R_{\mu\nu} \, , \hspace{1cm} L_2^{(1,2)} = \Box\phi\, R \, , 
\hspace{1cm} L_3^{(1,2)} = \phi^\mu \phi_{\mu\nu} \phi^{\nu}\,R \, , \nonumber \\  
&&L_4^{(1,2)} = \Box\phi\,\phi^{\mu}\phi^{\nu} R_{\mu\nu} \, , \hspace{1cm}
L_5^{(1,2)} =  \phi_{\alpha} \phi^{\alpha\beta} \phi_{\beta} \, \phi^{\mu}\phi^{\nu} R_{\mu\nu} \, ,
\ea
where we have used the notations $\phi_{\mu}=\nabla_\mu\phi$ and $\phi_{\mu\nu}=
\nabla_\mu\nabla_\nu\phi$. 

The mimetic constraint equation $\phi_\mu\phi^\mu=-1$ implies
the geodesic equation $\phi^\mu \phi_{\mu\nu}=0$ and therefore some of the higher derivative terms like 
$L_3^{(2,0)}$, $L_5^{(2,0)}$, $L_3^{(1,2)}$, $L_5^{(1,2)}$ will vanish if we impose the 
constraint $\phi^\mu \phi_{\mu\nu}=0$ in the on-shell action.  Therefore, one may think that these terms would not  contribute to the background and perturbed equations. But, the important point is that the
background and perturbed equations are obtained from the first and second orders actions respectively. Only after taking the variation we can impose the constraints $\phi_\mu\phi^\mu=-1$ and $\phi^\mu \phi_{\mu\nu}=0$. This is exactly the same  as in the mimetic term $\lambda(\phi_\mu\phi^\mu+1)$ in the action which vanishes if we naively impose the 
mimetic constraint $\phi_\mu\phi^\mu=-1$ from the start. 

The above classifications are based on the second derivative of $\phi$ such that the second 
derivative of mimetic field such as $\Box\phi$ is classified as the first order term while $(\Box\phi)^2$ 
and $\phi_{\mu\nu} \phi^{\mu\nu}$ are  second order and so on (see Refs. 
\cite{Achour:2016rkg,BenAchour:2016fzp,Langlois:2015skt,Langlois:2017mxy} for more details). 
Moreover, we have used the notation that the first upper index refers to the second derivative of the 
scalar field while the second index represents the degree associated with the curvature terms.

The curvature terms such as Ricci scalar $R$, Ricci tensor $R_{\mu\nu}$ and Riemann 
tensor $R_{\mu\nu\rho\sigma}$ are considered to be second order in second derivative of the scalar 
field. This can be understood if we work in uniform-$\phi$ slicing in which the future directed 
orthonormal vector to the hypersurfaces of constant $\phi$ is given by $n^{\mu} = \phi^\mu$. The
normalization condition $n_\mu n^{\mu}=-1$ is then equivalent to the mimetic constraint (\ref{mimetic-C}) 
and the extrinsic curvature is given by $K_{\mu\nu}=\nabla_\mu n_\nu=\phi_{\mu\nu}$. Therefore, in this
gauge, we can express all  terms in (\ref{L2i}) and (\ref{L3i}) in terms of purely geometric quantities. 
For example, the first order terms can be written as $\Box\phi=K^\mu_\mu=K$ and 
$\phi^\mu \phi_{\mu\nu} \phi^{\nu}=K_{\mu\nu} n^\mu n^\nu$. In the same manner, the second order 
terms can also be written as $\phi^{\mu\nu} \, R_{\mu\nu}= K^{\mu\nu} \, R_{\mu\nu}$, $\Box\phi\, R=K\, R$, 
and $\Box\phi\,\phi^{\mu}\phi^{\nu} R_{\mu\nu}=K\,R_{\mu\nu}n^{\mu}n^{\nu} $. We therefore can classify 
all terms in (\ref{L2i}) and (\ref{L3i}) just by considering the second derivatives of the scalar field. 

The potential term $V(\phi)$ in (\ref{action}) is required if one would like to  construct inflationary 
scenarios but we will show that it does not play any important role in stability analysis of the 
mode. The second term in (\ref{L2i}) is the well-known term which is added to the mimetic action 
(\ref{action0}) in \cite{Chamseddine:2014vna} in order to generate a constant sound speed for the 
cosmological perturbations.  Although the other  terms in (\ref{L2i}) generate nonzero sound speed, the corresponding cosmological perturbations suffer from ghost or gradient instabilities. As we have mentioned before, the 
terms like $f(\Box\phi) R$ could make the setup healthy \cite{Hirano:2017zox,Zheng:2017qfs} 
and therefore we have considered the cubic curvature dependent terms (\ref{L3i}). Naively  
these terms would generate Ostrogradsky ghosts in the metric sector of
the theory. But, as we will show, the mimetic constraint (\ref{mimetic-C}) prevents the propagation
of such ghosts. The cubic terms can also generate dynamical time dependent sound speed 
for the cosmological perturbations. One may ask what is the role of the quadratic terms 
(\ref{L2i}) when the cubic terms generate nonzero sound speed for the perturbations? We will
show that we need both of the quadratic terms (\ref{L2i}) and cubic curvature dependent terms (\ref{L3i}) 
in order to have healthy perturbations.

Apart from five cubic curvature dependent terms (\ref{L3i}), there are ten additional cubic
curvature independent terms as follows \cite{BenAchour:2016fzp}
\ba\label{L3i10}
&&L_1^{(3,0)} = (\Box\phi)^3 \, , \quad  L_2^{(3,0)} = \Box\phi \, \phi_{\mu\nu}
\phi^{\mu\nu}\, , \quad L_3^{(3,0)} = \phi_{\mu\nu} \phi^{\nu\alpha} 
\phi^{\mu}_{\alpha} \, ,  \quad L_4^{(3,0)} = (\Box\phi)^2 \phi^\mu \phi_{\mu\nu} 
\phi^{\nu}\, , \nonumber \\  
&&L_5^{(3,0)} = \Box\phi\, \phi^\mu \phi_{\mu\nu} \phi^{\nu\alpha}\phi_\alpha\, , 
 \quad L_6^{(3,0)} =  \phi_{\mu\nu} \phi^{\mu\nu} \phi_{\alpha} 
\phi^{\alpha\beta}\phi_{\beta} \, ,  \quad L_7^{(3,0)} = \phi_{\mu} \phi^{\mu\nu}
\phi_{\nu\alpha} \phi^{\alpha\beta} \phi_{\beta} \, , \nonumber \\  
&&L_8^{(3,0)} = \phi_{\mu} \phi^{\mu\nu} \phi_{\nu\alpha} \phi^{\alpha}
\phi_{\beta} \phi^{\beta\eta} \phi_\eta \, ,  \quad L_9^{(3,0)} = \Box\phi \,
(\phi^\mu \phi_{\mu\nu} \phi^{\nu})^2\, ,  \quad L_{10}^{(3,0)} = 
(\phi^\mu \phi_{\mu\nu} \phi^{\nu})^3 \, .
\ea
These terms however are not helpful for our purposes.  In Ref. \cite{Firouzjahi:2017txv} it is shown that the
mimetic setup in the presence of general function $f(\Box\phi)$ {\it i.e.} action (\ref{action1}),
suffers from ghost and gradient instabilities. Note that the first term in (\ref{L3i10}) is a special 
case of $f(\Box\phi)$. It is not difficult to show that the other terms in (\ref{L3i10}) also behave
like this term and therefore the cubic curvature independent terms in (\ref{L3i10}) cannot remove instabilities 
in mimetic setup.  We therefore neglect these terms throughout this paper.   

Moreover, we point out that we could also consider a non minimal coupling term
$f(\phi,X)\,R$. The mimetic constraint then implies that this coupling effectively is of the 
form of $g(\phi)\,R$ in which $g(\phi)=f(\phi,X)\mid_{X=-1}$ is only a function of $\phi$ 
which cannot play any significant role in our considerations. Therefore, we do not
consider this possibility either. 

%%%%%%%%%%%%%%%%%%%%%%%%%%%%%%%%%%%%%%%%%%%%%%%%%%%%%%%%%
\subsection{Cosmological background equations}

Now, let us calculate the background equations for our model (\ref{action}). 

The 0-0 
component of the  Einstein equations leads to the following modified Friedmann equation
\ba\label{Friedmann}
3H^2 &= & 
\frac{- 2 \bar{\lambda} + V - (6 \alpha_2-3\alpha_3) \dot{H}
 - 9 (\kappa_1+2 \kappa_2+4 \kappa_3-\kappa_4-\kappa_5) H^3}
{1 - 3(\alpha_1+ \alpha_2+ \alpha_3)} \\ 
& + & 3\, \frac{(4 \kappa_1+8 \kappa_2-14 \kappa_3+4 \kappa_4+5 \kappa_5) H \dot{H}
 + (\kappa_1+2 \kappa_2-2 \kappa_3-\kappa_4+\kappa_5) \ddot{H}}{1 - 3(\alpha_1+
\alpha_2+ \alpha_3)} \, , \nonumber
\ea
in which  we have imposed constraint  (\ref{phi}).

The i-i component of the Einstein equations gives the Raychuadhuri equation
\ba\label{V0}
2 \dot{H}+3 H^2 = \frac{V - 9 (\kappa_1+2 \kappa_2+\kappa_4) 
(H\dot{H} + H^3)}{1-\alpha_1-3 \alpha_2} \, .
\ea

Solving (\ref{V0}) for the potential $V(\phi)$ and 
then substituting the result into the Friedmann equation (\ref{Friedmann}), we find the following
solution for the auxiliary field
\ba\label{lambda0}
\bar{\lambda} & = & \frac{1}{2} \Big[ 3(2 \alpha_1+3 \alpha_3) H^2 + 
(2-2 \alpha_1-12 \alpha_2+3 \alpha_3) \dot{H} - 9 (4 \kappa_3-2 \kappa_4-
\kappa_5) H^3
\\ \nonumber
& + & 3 (7(\kappa_1+2 \kappa_2-2 \kappa_3+\kappa_4)+5 \kappa_5) H \dot{H} 
+ 3 (\kappa_1+2\kappa_2-2\kappa_3-\kappa_4+\kappa_5) \ddot{H} \Big] \, .
\ea

Varying the action with respect to $\phi$ yields the modified Klein-Gordon equation.

\ba\label{KGE}
&& 2\dot{\bar\lambda} + 6 H \bar{\lambda} - \dot{V}
+3 (2\alpha_2-\alpha_3)\ddot{H} - 3 (4 \alpha_1-6
\alpha_2+9\alpha_3) H \dot{H} - 9 (2 \alpha_1+3 \alpha_3)H^3
\nonumber \\
&& - 6 (2 \kappa_1+4 \kappa_2-7 \kappa_3) \dot{H}^2
- 18 (2 \kappa_1+ 9 \kappa_2 - 13\kappa_3) \dot{H} H^2 
-3 (7 \kappa_1+14 \kappa_2-20 \kappa_3) H \ddot{H} 
\nonumber \\
&& +108 \kappa_3H^4 -3 (\kappa_1+2\kappa_2-2 \kappa_3) \dddot{H}
-216 (\kappa_5-\kappa_4) \dot{H} H^3
-9 (5 \kappa_5-12\kappa_4) \dot{H} H^2 
\nonumber \\
&& + 3 (4\kappa_4-\kappa_5) H \dddot{H} + 9 (20 \kappa_4 - 9 \kappa_5)
H \dot{H}^2 + 9 (4 \kappa_4-\kappa_5) \dot{H} \ddot{H} - 81 \kappa_5 H^5 \,=\,0 \,,
\ea
in which we have substituted $\partial_{\phi}V=\dot{\phi}^{-1}\dot{V}=\dot{V}$. 

Only two of the three equations (\ref{Friedmann}), (\ref{V0}), and (\ref{KGE}) are independent. We use 
Eq. (\ref{V0}) and Eq. (\ref{lambda0}) to remove the potential $V(\phi)$ and $\bar{\lambda}$ in the follow up second order 
action of the  perturbations.

%%%%%%%%%%%%%%%%%%%%%%%%%%%%%%%%%%%%%%%%%%%%%%%%%%%%%

\subsection{Cosmological perturbations}
\label{comoving-sec}

In order to study the stability of the model in this section we present the analysis of cosmological 
perturbations in uniform-$\phi$ slicing. As we shall show, the analysis in this gauge are 
significantly easier than other gauges.  

We implement the  ADM formalism in which the metric is decomposed as
\begin{equation}\label{ADM-metric}
ds^2=-N^2dt^2+\gamma_{ij}(dx^i+N^i dt)(dx^j+N^j dt) \,,
\end{equation}
where $N$ and $N^i$ are the lapse function and the shift vector respectively while the 
three-dimensional metric $\gamma_{ij}$ determines the geometry of the spatial 
hypersurfaces. 

The action (\ref{action}) in terms of ADM variables takes the following form
\ba \label{actionADM}
S & = & \frac{1}{2}\int dt d^3x N \sqrt{\gamma} \left[\,{^{3}R} + N^{-2}(E_{ij}E^{ij}-E^2) \right] 
\nonumber \\
& + & \int dt d^3x N \sqrt{\gamma} \Big[\lambda(X+1) - V + \sum_{i=1}^5 \Big(\alpha_i 
L_i^{(2)} + \kappa_i L_i^{(3)}\Big) \Big]\,,
\ea
in which $^{3}R$ is the three-dimensional curvature associated with the spatial metric 
$\gamma_{ij}$, $E_{ij}$ is related to the Extrinsic curvature $K_{ij}$ via $E_{ij}=N\,K_{ij}$ which is given by
\begin{equation}\label{Edef}
E_{ij} = \frac{1}{2}(\dot{\gamma}_{ij} - \nabla_i N_j - \nabla_j N_i) \, ,
\end{equation}
and $E= E^i_i$ in which the spatial indices are raised and lowered by $\gamma_{ij}$.

%%%%%%%%%%%%%%%%%%%%%%%%%%%%%%%%%%%%%%%%%%%%%%%%%%%%%
\subsubsection{Scalar perturbations}

Here we study the scalar perturbations. 

There are four scalar degrees of freedom in 
metric (\ref{ADM-metric}) where one is given by the lapse function, another one by the shift vector and 
the two remaining ones are encoded in $\gamma_{ij}$. Moreover, there are two scalar degrees 
of freedom in matter sector associated with the scalar field $\phi$ and the auxiliary field $\lambda$. 
Thus, in total there are six scalar degrees of freedom in our setup. Two of these scalar degrees of 
freedom can be eliminated by fixing the gauge. In uniform-$\phi$ gauge, we eliminate one of the 
scalar degrees of freedom in matter part associated to $\phi$ as
\ba\label{deltaphi}
\delta\phi=0 \, ,
\ea 
as well as one scalar in $\gamma_{ij}$ such that the scalar perturbations in geometry are given by
\ba\label{ADM-scalar-pert-G}
N = 1+N_1 \, , \hspace{1cm} N^i = \partial^i \psi \, , \hspace{1cm} 
\gamma_{ij} = a^2 e^{2\calR}\delta_{ij} \,, 
\ea  
in which $\calR$ is the curvature perturbation. For the matter sector, there is 
only one degree of freedom associated to the auxiliary field $\lambda$ which is given 
by\footnote{Note that we are interested in the second order action and therefore we 
would generally consider the second order perturbations for the auxiliary field as $\lambda=
\bar \lambda+\lambda^{(1)}+\lambda^{(2)}$. But, the second order term $\lambda^{(2)}$ always 
multiplies with the zeroth order of mimetic constraint which vanishes since we work with 
the on-shell action subject to the mimetic constraint (\ref{mimetic-C}). However, one should consider the effect of this term when working with off-shell action (see section 5 of Ref. \cite{Firouzjahi:2017txv}).}
\ba\label{ADM-scalar-pert-M}
\lambda = \bar \lambda+\lambda^{(1)} \,, 
\ea
where $\bar{\lambda}$ denotes the background value of $\lambda$ while $\lambda^{(1)}$ is 
the first order scalar perturbation in $\lambda$.

Our next task is to substitute (\ref{ADM-scalar-pert-G}) and (\ref{ADM-scalar-pert-M}) in
(\ref{actionADM}) to obtain the second order action for the scalar perturbations. 
First, let us look at the 
mimetic constraint (\ref{mimetic-C}) at the level of perturbations. Substituting 
from (\ref{ADM-scalar-pert-G}) and (\ref{ADM-scalar-pert-M}) in (\ref{mimetic-C}), we find
\ba\label{alpha}
N_1 = 0 \,.
\ea
The above result simplifies our calculations considerably which is the reason why we have 
chosen to work in uniform-$\phi$ gauge.

The second order action for the scalar perturbations has a very complicated form (see section 
(\ref{appA-sec1}) of the appendix A for the details of calculations). However, it is straightforward 
to show that the quadratic action in Fourier space takes the following simple form

\ba\label{Lagrangian0}
S^{(2)} & = & \int d^3 k \int dt \, a^3 \Big[ - (1-\alpha_1-3 \alpha_2) (3\dot{\calR}_k 
+2 k^2 \psi_k) \dot{\calR}_k + (\alpha_1+\alpha_2) k^4 \psi_k^2
\nonumber \\
& + &  \frac{k^2}{a^2} \calR_k^2 - \frac{9}{2} (\kappa_1+2 \kappa_2+\kappa_4) H 
\dot{\calR}_k (2 k^2 \psi_k + 3 \dot{\calR}_k) - \frac{2}{a^2} (\kappa_1+2 \kappa_2) 
k^4\psi_k \calR_k
\nonumber \\
& - & \frac{1}{2} (9 \kappa_1+34 \kappa_2-13 \kappa_4) H k^4 \psi_k^2 \Big]\, .
\ea
Looking at the above action, we see that the coordinate $\psi_k$ has no time derivative. 
Therefore, we can eliminate $\psi_k$ by means of its equation of motion which is given by
\ba\label{EoM-psi}
&& -2 (\alpha_1+\alpha_2) k^2 \psi_k + 
(9 \kappa_1+34 \kappa_2-13\kappa_4) H k^2 \psi_k +2 (\kappa_1+2 \kappa_2) 
a^{-2} k^2 \calR_k
\nonumber \\
&&+ 2(1-\alpha_1-3 \alpha_2) \dot{\calR}_k - 9(\kappa_1+2 \kappa_2+\kappa_4) 
H \dot{\calR}_k = 0 \, .
\ea
Solving the above equation for $\psi_k$ gives
\ba\label{psi}
\psi_k = \frac{2 (\kappa_1+2 \kappa_2) a^{-2} \, \calR_k+ \big( 2 (1-\alpha_1-3 \alpha_2)
+9 (\kappa_1+2\kappa_2+\kappa_4)H \big) \,k^{-2}\dot{\calR}_k}{2 (\alpha_1+\alpha_2) 
- (9 \kappa_1+34 \kappa_2-13 \kappa_4) H}.
\ea

Note that in the above solution for $\psi$, the quadratic terms multiplied  by $\alpha_1$ and
$\alpha_2$ contribute only to the coefficient of $\dot{\calR}_k$ while the cubic curvature
dependent terms labeled by $\kappa_1$, $\kappa_2$, and $\kappa_4$ contribute not only
to the time derivative of curvature perturbation but also produce an extra term proportional 
to the spatial derivative of the curvature perturbation. The latter term is absent in the 
mimetic scenario without curvature dependent corrections which leads to the wrong sign of 
the gradient energy term. 
 
Substituting (\ref{psi}) into the Lagrangian (\ref{Lagrangian0}) we obtain the reduced second 
order action in term of curvature perturbation which, after some integration by parts,  gives 
the following simple  form
\ba\label{action2scalar}
S^{(2)} = \frac{1}{3} \int d^3 k \int dt \,
\Theta_s a^3\Big(\dot{\calR}_k^2-a^{-2}k^2c_s^2\calR_k^2\Big) \, ,
\ea
in which we have defined
\ba\label{Theta}
\frac{1}{\Theta_s}=\frac{1}{1+2 \alpha_1-3 (3 \kappa_1+14 \kappa_2-8 \kappa_4) 
H}-\frac{2}{2(1-\alpha_1-3 \alpha_2)+9 (\kappa_1+2 \kappa_2+\kappa_4) H} \, ,
\ea
while the sound speed of scalar perturbations is given by 
\ba\label{cs2}
c_s^2 & = & 3\Theta_s^{-1} \big[2 (\alpha_1+\alpha_2) - 
(9 \kappa_1+34\kappa_2-13\kappa_4) H\big]^{-2} 
\nonumber \\
& \times &
\Big[ 4 (\alpha_1+\alpha_2) \left(\kappa_1 (\alpha_1+3 \alpha_2+8)+2 \kappa_2
(\alpha_1+3 \alpha_2+16)-13 \kappa_4\right)\, H
\nonumber \\
& - & \Big(9 \kappa_1^2 (7+4 \alpha_1+8 \alpha_2)+4\kappa_1 \kappa_2 
(127+44 \alpha_1+96 \alpha_2)-4 \kappa_1 \kappa_4 (52+2 \alpha_1+15 \alpha_2)
\nonumber \\
& + & 4 \kappa_2^2 (255+52 \alpha_1+120\alpha_2)-8 \kappa_2 \kappa_4 
(104+2\alpha_1+15 \alpha_2)+169 \kappa_4^2\Big) \, H^2
\nonumber \\
& - & 2 (\kappa_1+2 \kappa_2) \big(9 \kappa_1 (1-2 \alpha_2)
+2 \kappa_2 (17-8 \alpha_1-42\alpha_2)-\kappa_4 
(13-22 \alpha_1-48 \alpha_2)\big) \dot{H}
\nonumber \\
& + & 9 (\kappa_1+2 \kappa_2) (9 \kappa_1+34\kappa_2-13 \kappa_4)
(\kappa_1+2\kappa_2+\kappa_4) \, H^3
\nonumber \\
& - & 4 (\alpha_1+\alpha_2)^2 + 2 a^{-2} k^2 (\kappa_1+2 \kappa_2)^2 
\big(2 (\alpha_1+\alpha_2)- (9 \kappa_1+34 \kappa_2-13 \kappa_4) H\big) \Big] \, .
\ea
Note that the term proportional to $ \calR_k$ in the r.h.s. of (\ref{psi}) contributes to the
gradient term of the effective Hamiltonian  ($k^2 \calR_k^2$) which makes it possible to have 
the correct sign for the sound speed and to avoid the gradient instability in our setup.

From the reduced Lagrangian (\ref{action2scalar}), the equation of motion for $\calR$ is
given by

\ba\label{EoMR}
\ddot{\calR}_k + \Big(3H+\frac{\dot{\Theta_s}}{\Theta_s}\Big) \dot{\calR}_k 
+ a^{-2}k^2c_s^2 \calR_k = 0 \, ,
\ea
which justifies  the definition (\ref{cs2}) as the sound speed.

%%%%%%%%%%%%%%%%%%%%%%%%%%%%%%%%%%%%%%%%%%%%%%%%%%%%%

\subsubsection{Tensor perturbations}
\label{Tensor-sec}
Now, we consider tensor perturbations in our model (\ref{action}). Tensor perturbations in 
the original mimetic dark matter scenario and also in the scenario with higher order quadratic 
term $(\Box\phi)^2$ are the same as in the standard general relativity \cite{Hirano:2017zox}. 
But it is shown that considering even the quadratic higher derivative term 
$\phi_{\mu\nu}\phi^{\mu\nu}$, which is the first quadratic term in (\ref{L2i}) in our setup, 
changes the kinetic part of the tensor perturbations \cite{Zheng:2017qfs}. Of more 
importance are the cubic curvature dependent terms (\ref{L3i}) which clearly would change 
the tensor perturbations. Therefore, in addition to the usual contributions from the  
Einstein-Hilbert term in the action, there would be some additional contributions for tensor 
perturbations in our setup.

Let us calculate the second order action for the tensor sector of our model (\ref{action}).
There are two tensor degrees of freedom in metric which are encoded in the spatial part 
of the metric as follows 
\ba\label{hij}
\gamma_{ij} = a^2 (\delta_{ij}+h_{ij}) \, ,
\ea
where $h_{ij}$ represents the tensor perturbations which satisfies the traceless and transverse
conditions $h_{ii}=0$ and $\partial_i h_{ij}=0$. Without loss of generality, we choose a 
coordinate system in which the momentum ${\bf k}$ is oriented along the three axis. The 
traceless and transverse conditions then imply $h_{22}=-h_{11}$ and $h_{i3}=0$. In this 
respect, the two tensor modes can be completely characterized by the two components
$h_{11}=h^{+}$ and $h_{12}=h^{\times}$. We consider the Fourier transformation 
\ba\label{hijFT}
h_{ij}=\sum_{s=+,\times}\int \frac{d^3k}{(2\pi)^3}
\epsilon^s_{ij}(k)\,h^s_k(t) e^{i{\bf k}.{\bf x}} \, ,
\ea
in which $\epsilon_{ij}$ represents the polarizations base of the gravitational waves which satisfies the conditions 
$\epsilon_{ii}=0=k_i\epsilon_{ij}$ and $\epsilon^s_{ij}(k)\epsilon^{s'}_{ij}(k)=2\delta_{s{s'}}$. 
Substituting (\ref{hijFT}) in action (\ref{actionADM}) and using the background equation (\ref{V0}), 
it is straightforward to show that the second order action for tensor perturbations will be
\ba\label{action2tensor}
S^{(2)} =  \frac{1}{4}\sum_{s=+,\times} \int d^3 k \int dt \,
\Theta_t a^3 \Big(\dot{h}^s_k\dot{h}^s_k- a^{-2}k^2c_t^2 h^s_kh^s_k\Big) \, ,
\ea
where we have defined
\ba\label{ThetaT}
\Theta_t = 1+2\alpha_1 - 3 (\kappa_1+2 \kappa_2-2\kappa_4) H \,,
\ea
and the gravitational wave speed as
\ba\label{ct2}
c_t^2 = \Theta_t^{-1}\big(1-3 (\kappa_1+2\kappa_2) H \big) \, .
\ea

From the action (\ref{action2tensor}), the equation of motion for two tensor modes are given by

\ba\label{EoMh}
\sum_{s=+,\times} \ddot{h}^s_k + \Big(3H+\frac{\dot{\Theta_t}}{\Theta_t}\Big) 
\dot{h}^s_k + a^{-2}k^2c_t^2 h^s_k  = 0 \, .
\ea
As expected, the cubic
curvature terms (\ref{L3i}) change the tensor perturbations such that the speed of
gravitational wave (\ref{ct2}) is different than unity. 

%%%%%%%%%%%%%%%%%%%%%%%%%%%%%%%%%%%%%%%%%%%%%%%%%%%%%
\subsection{Stability conditions}

Looking at the second order action for the scalar perturbations (\ref{action2scalar}) and
the second order action for the tensor perturbations (\ref{action2tensor}), one immediately
finds that the model (\ref{action}) is stable if the following conditions are  satisfied
\ba\label{StabilityConditions}
c_s^2>0 \,,\hspace{1cm} \Theta_s>0 \,,\hspace{1cm}
c_t^2>0\,,\hspace{1cm} \Theta_t>0 \,.
\ea

These conditions would be satisfied by appropriate choices of the parameters $\alpha_1$,
$\alpha_2$, $\kappa_1$, $\kappa_2$, and $\kappa_4$. It is important to note that the existence 
of quadratic terms (\ref{L2i}) is essential in order to remove the instabilities. This can be 
seen from the effects of these terms on the kinetic parts of the scalar and
tensor perturbations, Eqs.  (\ref{Theta}) and (\ref{ThetaT}). So, we need to have both the quadratic curvature independent
terms (\ref{L2i}) and the cubic curvature dependent terms (\ref{L3i}) in order for the cosmological perturbations to be stable.

%%%%%%%%%%%%%%%%%%%%%%%%%%%%%%%%%%%%%%%%%%%%%%%%%%%%%
\section{Quintic mimetic model}

In the previous section, we have found stable extension of the mimetic gravity by adding quadratic 
curvature independent terms (\ref{L2i}) together with the cubic curvature dependent terms 
(\ref{L3i}) to the original mimetic model (\ref{action0}). This result was also  reported 
in Refs. \cite{Hirano:2017zox,Zheng:2017qfs} while here we extend their results by considering all the relevant terms. 
We have shown that, in comparison with the Horndeski theories, it is not necessary to fix the coefficients of 
the higher derivative terms in mimetic scenario. This point would help us to control the 
background equations and therefore to find a healthy mimetic dark matter and healthy mimetic
nonsingular scenario. 

To be more precise, in original model of mimetic gravity
\cite{Chamseddine:2013kea}, the field $\phi$ mimics the roles of dark matter particles. However, as we have seen in the previous section,  adding a coupling between the second derivative of $\phi$ and curvature such as $f(\Box\phi) R$ 
can make the setup stable  but the corresponding background equations, such as obtained in Eqs.  (\ref{Friedmann}) and (\ref{V0}),  are significantly different from those of  Eqs. (\ref{Friedmann-GHD}) and (\ref{Raychuadhuri-GHD})  and the mimetic field may not play the roles of dark matter. In addition, it is 
not clear how a nonsingular bouncing mimetic scenario such as in \cite{Chamseddine:2016uef} can be 
realized when we add a coupling between the second derivative of $\phi$ and curvature. 
%Working  with the classified higher derivative terms like (\ref{L2i}) and (\ref{L3i}) allows us to control both  the background and the perturbed equations to find healthy extension of  mimetic dark matter and mimetic nonsingular scenarios.
%Note that both the mimetic dark matter and mimetic nonsingular models are special cases of  the model (\ref{action1}) with different functional forms for $f(\Box\phi)$. However,  the model  (\ref{action1}) suffers from some linear instabilities independent of the form of $f(\Box\phi)$ \cite{Firouzjahi:2017txv}. 
In this section, working with the classified higher derivative terms,  
our aim is to find the healthy extension of the 
model (\ref{action1})  such that the background  equations continue to take the simple forms of   Eqs. (\ref{Friedmann-GHD}) and  (\ref{Raychuadhuri-GHD}), allowing us to construct  healthy extensions of the 
mimetic dark matter and mimetic nonsingular bounce scenarios.

The first choice is to consider cubic curvature dependent terms (\ref{L3i}) in the model (\ref{action1}) and 
then arrange the coefficients such that the background equations keep the forms of 
Eqs. (\ref{Friedmann-GHD}) and  (\ref{Raychuadhuri-GHD}). In 
appendix B, we have shown that it is not possible to find a healthy extension of the model (\ref{action1}) 
such that the background equations have the forms of  Eqs. (\ref{Friedmann-GHD}) and 
(\ref{Raychuadhuri-GHD}). The next choice is to consider the quartic curvature dependent terms such as 
$(\Box\phi)^2 R$.  Our analysis show that these term also cannot make the model (\ref{action1}) 
stable for general functional form of $f(\Box\phi)$ if we demand the background equations to remain unchanged (see the appendix B
for the details). For instance, our analysis show that for the particular choice of $f(\Box\phi)$ employed 
in the nonsingular mimetic model of \cite{Chamseddine:2016uef}, as given by Eq. (\ref{f}) below,  the quartic terms cannot make  the setup stable. In this view, we add the quintic curvature dependent terms to the action 
(\ref{action1}) as follows 
\ba\label{action-Bounce-Healthy}
S= \int d^4 x  \sqrt{-g} \Big[\, \frac{1}{2} R + \lambda \left(X+ 1 \right)+ f(\chi) 
+ \sum_{i=1}^{22}\,\delta_i L^{(3,2)}_i + \sum_{i=1}^{12}\,\beta_i L^{(1,4)}_i \,\Big] \, ,
\ea
in which $\delta_i$ and $\beta_i$ are the coefficients of the quintic curvature dependent terms 
$L^{(3,2)}_i$ and $L^{(1,4)}_i$ which are defined as
\ba\label{L5i32}
&&L_1^{(3,2)} = (\Box\phi)^3\,R \, , \hspace{2.2cm} L_2^{(3,2)} = \Box\phi 
\, \phi_{\mu\nu} \phi^{\mu\nu} \,R\, , \hspace{2cm} 
L_3^{(3,2)} = \phi_{\mu\nu} \phi^{\nu\alpha} \phi^{\mu}_{\alpha} \, R \, , 
\nonumber \\
&& L_4^{(3,2)} = (\Box\phi)^2 \phi^\mu \phi_{\mu\nu} \phi^{\nu} \, R \, ,  \hspace{.7cm}
L_5^{(3,2)} = \Box\phi\, \phi^\mu \phi_{\mu\nu} \phi^{\nu\alpha}\phi_\alpha \, R \, , 
\hspace{1.1cm} L_6^{(3,2)} =  \phi_{\mu\nu} \phi^{\mu\nu} \phi_{\alpha} 
\phi^{\alpha\beta}\phi_{\beta} \, R \, , \nonumber \\
&&L_7^{(3)} =  \phi_{\mu} \phi^{\mu\nu}\phi_{\nu\alpha} \phi^{\alpha\beta} \phi_{\beta} 
\, R ,  \hspace{.8cm} L_8^{(3,2)} = \phi_{\mu} \phi^{\mu\nu} \phi_{\nu\alpha} \phi^{\alpha}
\phi_{\beta} \phi^{\beta\eta} \phi_\eta R \, , \hspace{.3cm} L_9^{(3,2)} = \Box\phi \,
(\phi^\mu \phi_{\mu\nu} \phi^{\nu})^2\, R \, , \nonumber \\
&&L_{10}^{(3,2)} = (\phi^\mu \phi_{\mu\nu} \phi^{\nu})^3 \, R \, , \hspace{1.3cm}
L_{11}^{(3,2)} = (\Box\phi)^3\,\phi^{\mu}\phi^{\nu} R_{\mu\nu} \, , \hspace{1.5cm} 
L_{12}^{(3,2)} = \Box\phi \, \phi_{\mu\nu} \phi^{\mu\nu} \,\phi^{\alpha}\phi^{\beta} R_{\alpha\beta}\, ,
\nonumber \\
&&L_{13}^{(3,2)} = \phi_{\mu\nu} \phi^{\nu\alpha} \phi^{\mu}_{\alpha} \, 
\phi^{\sigma}\phi^{\rho} R_{\sigma\rho} \, ,   \hspace{2.1cm}
L_{14}^{(3,2)} = (\Box\phi)^2 \phi^\mu \phi_{\mu\nu} \phi^{\nu} \, 
\phi^{\sigma}\phi^{\rho} R_{\sigma\rho} \, ,  \nonumber \\
&&L_{15}^{(3,2)} = \Box\phi\, \phi^\mu \phi_{\mu\nu} \phi^{\nu\alpha}\phi_\alpha \,
\phi^{\sigma}\phi^{\rho} R_{\sigma\rho} \, , 
\hspace{1cm} L_{16}^{(3,2)} =  \phi_{\mu\nu} \phi^{\mu\nu} \phi_{\alpha} 
\phi^{\alpha\beta}\phi_{\beta} \, \phi^{\sigma}\phi^{\rho} R_{\sigma\rho} \, , \nonumber \\
&&L_{17}^{(3)} =  \phi_{\mu} \phi^{\mu\nu}\phi_{\nu\alpha} \phi^{\alpha\beta} \phi_{\beta} 
\, \phi^{\sigma}\phi^{\rho} R_{\sigma\rho} \, ,  
\hspace{1.2cm} L_{18}^{(3,2)} = \phi_{\mu} \phi^{\mu\nu} \phi_{\nu\alpha} \phi^{\alpha}
\phi_{\beta} \phi^{\beta\eta} \phi_\eta \, \phi^{\sigma}\phi^{\rho} R_{\sigma\rho} \, , 
\nonumber \\
&&L_{19}^{(3,2)} = \Box\phi \,
(\phi^\mu \phi_{\mu\nu} \phi^{\nu})^2\, \phi^{\sigma}\phi^{\rho} R_{\sigma\rho}\, ,  \hspace{1.1cm}
L_{20}^{(3,2)} = (\phi^\mu \phi_{\mu\nu} \phi^{\nu})^3 \, \phi^{\sigma}\phi^{\rho} R_{\sigma\rho} \, ,
\nonumber \\
&&L_{21}^{(3,2)} = \Box\phi\, \phi^{\alpha\mu} \phi^{\beta\nu} R_{\mu\nu\alpha\beta} 
\, ,  \hspace{2.4cm}
L_{22}^{(3,2)} = \phi^\mu \phi_{\mu\nu} \phi^{\nu}\, \phi^{\alpha\eta} \phi^{\beta\sigma}
R_{\eta\sigma\alpha\beta}  \, ,
\ea
and 
\ba\label{L5i14}
&& L_1^{(1,4)} = \phi^{\mu\nu} R_{\mu\nu} R \, , \hspace{1.75cm} L_2^{(1,4)} = \Box\phi \, 
R^2\, , \hspace{3cm} L_3^{(1,4)} = \phi^\mu \phi_{\mu\nu} \phi^{\nu} \,R^2\, , \\ 
&& L_4^{(1,4)} =  \Box\phi \, \phi^{\mu}\phi^{\nu}R_{\mu\nu} R\, , \hspace{.8cm}
L_5^{(1,4)} =  \phi_{\alpha} \phi^{\alpha\beta} \phi_{\beta} \phi^{\mu}\phi^{\nu}R_{\mu\nu} 
R\, , \hspace{.8cm} L_6^{(1,4)} = \phi^{\alpha\beta} R_{\alpha\beta}\, \phi^{\mu}\phi^{\nu}
R_{\mu\nu} \, ,\nonumber \\ 
&& L_7^{(1,4)} =  \Box\phi \, (\phi^{\mu}\phi^{\nu}R_{\mu\nu})^2\, , \hspace{0.7cm}
L_{8}^{(1,4)} = \phi_{\alpha} \phi^{\alpha\beta} \phi_{\beta} (\phi^{\mu}\phi^{\nu} 
R_{\mu\nu})^2\, , \hspace{0.6cm} L_{9}^{(1,4)} = \Box\phi \, R_{\mu\nu} R^{\mu\nu} \,, 
\nonumber \\
&& L_{10}^{(1,4)} =  \phi_{\alpha} \phi^{\alpha\beta} \phi_{\beta} R_{\mu\nu} R^{\mu\nu} 
\,, \hspace{.5cm} L_{11}^{(1,4)} = \Box\phi \, R_{\mu\nu\eta\sigma} R^{\mu\nu\eta\sigma} 
\hspace{1.7cm} L_{12}^{(1,4)} =  \phi_{\alpha} \phi^{\alpha\beta} \phi_{\beta} 
R_{\mu\nu\eta\sigma} R^{\mu\nu\eta\sigma} \, . \nonumber  
\ea
The above are the independent quintic terms which we have found so far but there may still be other 
independent quintic terms. However, the above terms are sufficient for our purpose in this work.

There are two types of quintic curvature dependent terms constructed from coupling between 
the higher derivatives of the scalar field and the curvature terms. In cubic terms (\ref{L3i}) and quartic terms 
(\ref{L4i}), the degree of the curvature terms, which is labeled by the second upper index, was two. 
It can be however raised to four in quintic model (\ref{L5i14}). The terms (\ref{L5i14}) would 
generally produce Ostrogradsky ghost in the metric sector of the model and therefore one may 
neglect them and work only with (\ref{L5i32}). But, we can choose the coefficients of these 
terms such that the setup to be free of Ostrogradsky ghost.  We will show that the appearance 
of these terms in quintic model makes it very different from the quartic model such that the 
instabilities can be removed in this model for general functional form of $f(\Box\phi)$.

%%%%%%%%%%%%%%%%%%%%%%%%%%%%%%%%%%%%%%%%%%%%%%%%%%%%%
\subsection{Background equations}

Our first task is to find the background equations associated with the model 
(\ref{action-Bounce-Healthy}). The 0-0 component of the Einstein equations  leads to the following modified Friedmann equation 
\ba\label{Friedmann-f}
3H^2 &=& -2 {\bar\lambda} -( f + 3H f_{\chi} + 3 \dot{H} f_{\chi\chi} )
+ 9  \Delta_2 H^5 + 9 \Delta_3 ( 2H \dot{H}^2 + H^2 \ddot{H})
+ 3 \Delta_5 \dot{H} H^3
\nonumber \\
&+& 9 B_4 H^5 - 3 B_5 H^2 \ddot{H} - 3 B_6 H \dot{H}^2 
- 6 B_7 \dot{H} H^3 - 6 B_8 \ddot{H}^2 \,,
\ea
while  the i-i components give the  Raychuadhuri equation as follows 
\ba\label{Raychuadhuri-f}
2 \dot{H}+3H^2 & = & - (f+ 3H f_{\chi} -3 \dot{H}  f_{\chi\chi}) 
- 3 \Delta_1( 5\dot{H} +  3 H^2) H^3 \nonumber \\ 
& - & 6 B_1 \big( 6 H^2 \ddot{H} + 9 H \dot{H}^2+ 2 \dot{H} \ddot{H} 
+ H \dddot{H}\, \big) + 18 B_2 H^5 + 6 B_3 \dot{H} H^3 \,, 
\ea
in which the coefficients $B_i$ and $\Delta_i$ denote the different combinations of the 
coefficients $\beta_i$ and $\delta_i$ respectively which are shown in section (\ref{appA-sec2}) 
of the appendix A.

In the above background equations, the contributions from the curvature dependent terms 
(\ref{L5i32}) and (\ref{L5i14}) are labeled by new coefficients $\Delta_i$ and $B_i$ respectively. 
We therefore demand that $B_i=0$ and $\Delta_i=0$ to ensure that the curvature dependent terms 
(\ref{L5i32}) and (\ref{L5i14}) do not change the background equations. Indeed, all of the 
combinations (\ref{Bi}) and (\ref{deltai}) are not independent and it is sufficient to fix only seven 
coefficients of $\beta_i$ and $\delta_i$. Without loss of generality, we fix $\beta_{12}$, 
$\beta_{11}$, and $\beta_{7}$,  $\delta_{22}$, $\delta_{21}$, $\delta_{16}$, and 
$\delta_3$ as follows:
\ba\label{Sol-beta-delta}
&& \beta_7 = \frac{1}{6} (6 \beta_4+\beta_6) \, , \nonumber \\
&&\beta_{12}= \frac{1}{4} \big( 4 \beta_1-4 \beta_{10}-12 \beta_3+6 \beta_5-2 \beta_6
-3 \beta_8 \big) \, ,  \nonumber\\
&&\beta_{11}= \frac{1}{8} \big(-4 \beta_1-24 \beta_2+6 \beta_4+\beta_6-8 \beta_9 \big) \, , \nonumber\\
%&& \beta_7 = \frac{1}{6} (6 \beta_4+\beta_6) \, , \nonumber \\
&& \delta_{22} = \frac{1}{4} \big(-110 \beta_1+8 \beta_{10}-216 \beta_2+48 \beta_3-27 \beta_4-24 \beta_5+\beta_6+6 \beta_8-36 \beta_9 \nonumber \\
&&+54 \delta_1+9 \delta_{11}+3\delta_{12}+\delta_{13}+6 \delta_2-18\delta_4-6 \delta_6 \big) \, ,  \nonumber \\
&&
\delta_{21} = \frac{1}{4} \big(-14 \beta_1-24 \beta_2-9\beta_4-\beta_6-4 \beta_9-5 (9 \delta_{11}+3\delta_{12}+\delta_{13})\big) \, ,  \nonumber \\
&&
\delta_{16} = \frac{1}{12} \big(-30 \beta_1-8 \beta_{10}-24
\beta_2-48 \beta_3-39 \beta_4+12 \beta_5-3 \beta_6-4\beta_9 \nonumber \\ 
&&-3 (18 \delta_1+9 \delta_{11}+11\delta_{12}+5 \delta_{13}+12 \delta_{14}+2\delta_2
-6 \delta_4-2 \delta_6)\big) \, ,  \nonumber \\
&&
\delta_3 = - \big(14 \beta_1 + 24 \beta_2+12 \beta_4+2\beta_6 +4 \beta_9+9 \delta_1
+3 \delta_2 \big) \, .  
\ea

Substituting the above solutions in (\ref{Friedmann-f}) and (\ref{Raychuadhuri-f}), we obtain the
desired background equations  in the forms of Eqs. (\ref{Friedmann-GHD}) and 
(\ref{Raychuadhuri-GHD}). Therefore, the quintic model (\ref{action-Bounce-Healthy}) and the original model 
(\ref{action1}) lead to the same background equations when conditions (\ref{Sol-beta-delta})
are met. At the level of the perturbations however these two models are very different. As shown in \cite{Firouzjahi:2017txv} the  model (\ref{action1}) suffers from some linear instabilities while, as we 
shall show in the next subsection, the quintic model (\ref{action-Bounce-Healthy}) can have 
healthy perturbations. 

%%%%%%%%%%%%%%%%%%%%%%%%%%%%%%%%%%%%%%%%%%%%%%%%%%%%%
\subsection{Stability analysis: Scalar and tensor perturbations}

We  do not repeat the details of calculations here. Following the same steps as in previous section, and with tedious but straightforward calculations, one can show that the second order action for the scalar perturbations in Fourier space will be
\ba\label{action2scalar-beta}
S^{(2)} &=& \frac{1}{2} \int d^3 k \int dt\,
a^3 \bigg[ (3f_{\chi\chi}-2) (3\dot{\calR}_k^2 + 2 k^2 \psi_k \dot{\calR}_k) 
+ 2 a^{-2}k^2 \calR_k^2 + f_{\chi\chi} k^4 \psi_k^2 
\\
&+& \frac{3}{2}B'_1a^{-2}Hk^2 \dot{\calR}_{k}^{2} 
+4 a^{-2} H k^4 B'_{2}{{\dot{\calR}}_{k}}{{\psi }_{k}}
+ B'_3 H k^4 \dot{\psi }_{k}^{2}
\nonumber \\
&-& \Big(2 B'_4 a^{-2} k^2
+ 3 B'_{5} (3 f_{\chi \chi } - 2)^{-1} (3f+ 9H{f_{\chi }}+3H^2{{f}_{\chi \chi }}
+7H^2) \Big) a^{-2} H k^2 \calR_{k}^{2}\nonumber\\
&+& \frac{2a^{-2}k^4}{ 3{{f}_{\chi \chi }} - 2 } \big( B'_6 (f + 3 H {{f}_{\chi }}) 
- H\left( B'_7 - 3 B'_{8} {f_{\chi \chi }} \right) \big) {{\calR}_{k}}{{\psi }_{k}}\nonumber\\
&+& \Bigl( 2B'_{9} H (3{{f}_{\chi \chi }}-2 ) \left( f-3H{{f}_{\chi }} \right)
+\left( 3B'_{10}{f_{\chi \chi }}-2B'_{11} \right) H^3 \nonumber\\
&-&\frac{9B'_1{f_{\chi \chi \chi }}}{{3{{f}_{\chi \chi }}-2}}
\bigl({{f}^{2}}+6fH\left( {{f}_{\chi }}+H \right)+9{{H}^{2}}f (f_{\chi }^{2}+{{H}^{2}})
-18{{H}^{3}}{{f}_{\chi }} \bigl)\Bigl) k^4 \psi _{k}^{2}\bigg] \nonumber ,
\ea
where the coefficients $B'_i$ denote different combinations of $\beta_i$ and $\delta_i$ which
are shown in section (\ref{appA-sec2}) of the appendix A.

In deriving (\ref{action2scalar-beta}) we have imposed the conditions (\ref{Sol-beta-delta}) which
allow us to substitute ${\bar{\lambda}}$ and $\dot{H}$ from the background equations
(\ref{Friedmann-GHD}) and (\ref{Raychuadhuri-GHD}). 

In the model (\ref{action}), the variable $\psi_k$ has appeared with no time derivative in the corresponding 
second order action (\ref{Lagrangian0}). However, in the second order action (\ref{action2scalar-beta}), 
there is  a term proportional to $\dot{\psi}_k^2$ which cannot be eliminated via integration by parts. The 
appearance of this term signals the propagation of a new degree of freedom in the model 
(\ref{action-Bounce-Healthy}) which is nothing but the so-called Ostrogradsky ghost. Looking at the 
coefficients of this term, we  see that this term comes from the quintic terms (\ref{L5i14}) that are quadratic 
in curvature terms. In order to remove this unstable mode, we demand that the coefficient of this term to 
be zero, yielding  $B'_3=4{{\beta }_{1}}-{{\beta }_{6}}+3\left( 2{{\beta }_{9}}
+8{{\beta }_{2}} \right) =0$. Solving for $\beta_9$ from this equation gives
\ba\label{Solb9}
\beta_9= \frac{1}{6} (-4 \beta_1 - 24 \beta_2 + \beta_6) \, .
\ea
Substituting the above solution into the action (\ref{action2scalar-beta}), $\psi_k$ appears with no time 
derivative and therefore it can be removed by its equation of motion
\footnote{We note that here we have just removed the ghosts at the level of linear
perturbations since we were interested in linear stability of the model. 
In order to study the stability of the model to all orders, one has to perform a nonlinear analysis and it is not clear 
whether or not our condition (\ref{Solb9}) is sufficient.}. The resultant reduced actionin 
term of curvature perturbation $\calR_k$ takes a very complicated form. To study the 
stability of the model, 
however, it is not necessary to obtain the explicit form of the action. The first line in 
(\ref{action2scalar-beta}) are the terms which are in common with the action (\ref{action1}) while the other 
lines which are labeled by $B'_i$ come from the quintic curvature dependent terms (\ref{L5i32}) and 
(\ref{L5i14}). We know that, in the absence of the curvature dependent terms, the model suffers from ghost
and gradient instabilities \cite{Firouzjahi:2017txv}. Moreover, the term $\psi_k \calR_k$ in the fourth line 
shows that the solution for $\psi_k$ gives contribution to the gradient term $\calR_k^2$ and therefore can 
remove the gradient instability in the same manner that the cubic terms (\ref{L3i}) remove the gradient 
instability in model (\ref{action}). In appendix B, we have shown that this term is absent in the cubic model 
when we demand that the background equations remain unchanged. This term however is present in 
quartic model and one may ask what is the advantage of the quintic model in comparison with the quartic 
model. Indeed, the appearance of the term proportional to $\psi_k {\dot\calR}_k$ in the second line of 
(\ref{action2scalar-beta}) makes the quintic model very different from quartic model since this term is 
absent in quartic model (\ref{action2scalar-theta}). In this respect, independent of the form of $f(\Box \phi)$, 
we can separately control both the kinetic term and the gradient term. As a result we expect  that the quintic 
model (\ref{action-Bounce-Healthy}) can have healthy scalar perturbations independent of the form of 
$f(\Box \phi)$.

The coefficient of ${\dot\psi}_k^2$ in (\ref{action2scalar-beta}) (which is given by $B'_3$ 
defined in Eq. (\ref{Bprimes}) in Appendix \ref{appA})  is originated from the terms $L_{i}^{(1,4)}$ defined in (\ref{L5i14}) which  include the quadratic terms in curvatures. In standard general relativity, it is well known that the purely quadratic  curvature terms $R_{\mu\nu}R^{\mu\nu}$ and $R_{\mu\nu\rho\sigma}R^{\mu\nu\rho\sigma}$ produce  Ostrogradsky ghost. In mimetic gravity the situation is even worse such that the $f(R)$ extension of  mimetic gravity \cite{Nojiri:2014zqa} suffers from Ostrogradsky ghost \cite{Hirano:2017zox} while the $f(R)$  model in standard general relativity is free of Ostrogradsky ghost. So, in addition to $R_{\mu\nu}R^{\mu\nu}$  and $R_{\mu\nu\rho\sigma}R^{\mu\nu\rho\sigma}$, the terms $R^2$ would produce Ostrogradsky ghost in mimetic gravity. The question then arises: is it possible to find a healthy extension of mimetic scenario from purely gravitational higher derivative term 
$f(R,R_{\mu\nu}R^{\mu\nu},R_{\mu\nu\rho\sigma}R^{\mu\nu\rho\sigma})$? We therefore consider 
the simplest case of quadratic term and add the term
$c_1 R^2+c_2 R_{\mu\nu}R^{\mu\nu}+c_3R_{\mu\nu\rho\sigma}R^{\mu\nu\rho\sigma}$ to the original mimetic action (\ref{action0}) where $c_i$ are arbitrary constants. We now apply our method to this scenario {\it i.e.} manipulating the coefficients $c_i$ such that the model be free of Ostrogradsky ghost. 
Performing the calculations, one finds that the necessary condition to avoid the Ostrogradsky ghost is 
$c_1=-\frac{1}{4}c_2=c_3$ which shows that the only possibility is the Gauss-Bonnet term 
$c_1\left(R^2-4R_{\mu\nu}R^{\mu\nu}+R_{\mu\nu\rho\sigma}R^{\mu\nu\rho\sigma}\right)$. The 
Gauss-Bonnet term does not contribute to the background equations and therefore the setup describes 
dark matter, much similar to the original setup (\ref{action0}). At the level of perturbations, there is just 
one independent parameter, which without loss of generality, we choose it to be $c_1$. The results show 
that the gradient term remain unaffected such that it again appears with a wrong sign while the ghost 
instability in kinetic term can be avoided for $c_1<0$ and $H^2< M_{{\rm Pl}}^2/48 | c_1|$.

Now let us study the tensor perturbations. Following the same steps as in previous section, it is 
straightforward to obtain the second order action for the tensor perturbations as follows 
\ba\label{action2tensor-beta}
S^{(2)} = \frac{1}{4} \sum_{s=+,\times} \int d^3 k \int dt \,
a^3 \Theta_t \Big(\dot{h}^s_k\dot{h}^s_k-a^{-2}k^2c_t^2 h^s_kh^s_k\Big) \, ,
\ea
in which we have defined 
\ba\label{Thethat-beta}
\Theta_t &=& 1+ 27 B'_1 {f}_{\chi \chi \chi }(3{{f}_{\chi \chi }}-2)^{-3}
\big( {{f}^{2}}+3{{f}_{\chi }}-6fH\left( H+{{f}_{\chi }} \right) \big)\nonumber\\
&-& H (3{{f}_{\chi \chi }}-2)^{-1}\Big(6 B'_{12} H{{f}_{\chi }}+2B'_{15}f-
\left( 2B'_{13}+3B'_{14} f_{\chi\chi} + 486 B'_1 f_{\chi}f_{\chi\chi\chi} \right) H^2 \Big),
\ea
and the speed of gravitational wave as
\ba\label{cs2-beta}
c_t^2&=&\Theta_t^{-1}\bigg[ 1 + \frac{1}{2} B'_{15} a^{-2} H k^2 
- \frac{\left( B'_{16}+3B'_{17} f_{\chi \chi } \right) H^3
+ B'_{18} (f+3 f_{\chi} H) H}{3f_{\chi \chi } -2} \nonumber\\
&-&18 B'_{19}\frac{f_{\chi\chi\chi}\left(f^2+9{{H}^{4}}+18{{H}^{3}}
+9f_{\chi}^2H^2+6H{{f}_{\chi }}
\left( H+{{f}_{\chi }} \right) \right)}{{{(3f_{\chi\chi} -2 )}^{3}}} \bigg] \, ,
\ea
with $B'_i$ are shown in section (\ref{appA-sec2}) in the appendix A.

In deriving action (\ref{action2tensor-beta}), we have used background equations
(\ref{Friedmann-GHD}) and (\ref{Raychuadhuri-GHD}) and also we have imposed the 
conditions (\ref{Sol-beta-delta}) and (\ref{Solb9}).

Looking at the action (\ref{action2tensor-beta}), we see that the model 
(\ref{action-Bounce-Healthy}) can have healthy tensor perturbations. Indeed, the action
(\ref{action}) was already providing healthy tensor perturbations. The important difference 
between our quintic model (\ref{action-Bounce-Healthy}) and the model (\ref{action}) is that the 
speed of gravitational wave in our setup depends on the wave number and is different than unity.

In summary, our analysis show that the quintic model (\ref{action-Bounce-Healthy}) 
can potentially produce healthy perturbations for both scalar and tensor perturbations while  the background equations remain unchanged as in Eqs. (\ref{Friedmann-GHD}) and (\ref{Raychuadhuri-GHD}). 
%But,  the explicit confirmation of this claim requires to find specific values for all parameters of the  model which is not an easy task due to the large number of parameters involved.

%%%%%%%%%%%%%%%%%%%%%%%%%%%%%%%%%%%%%%%%%%%%%%%%%%%%%
\section{Applications of higher derivative mimetic model}

Let us now remark some applications of the quintic model which we have introduced in the 
previous section.

%%%%%%%%%%%%%%%%%%%%%%%%%%%%%%%%%%%%%%%%%%%%%%%%%%%%%
\subsection{Mimetic dark matter}
One of the most interesting feature of mimetic gravity is that the mimetic
field can play the roles of dark matter \cite{Chamseddine:2013kea,Mirzagholi:2014ifa}. However, 
the corresponding perturbations have zero sound speed and therefore the quadratic higher 
derivative term $(\Box\phi)^2$ was added to the action to generate  nonzero sound speed 
\cite{Chamseddine:2014vna} (see however \cite{DeFelice:2015moy, Gumrukcuoglu:2016jbh, Babichev:2016jzg, Babichev:2017lrx} where it is argued that having a zero sound speed is not a problem). 
Interestingly, the setup again describes dark matter at the 
background level \cite{Mirzagholi:2014ifa} but it now suffers from some linear instabilities 
\cite{Ramazanov:2016xhp,Ijjas:2016pad}. As we discussed in the Introduction section, these 
instabilities can be removed by considering a coupling between the second derivative of the 
scalar field and curvature terms such as $\Box\phi R$ \cite{Hirano:2017zox,Zheng:2017qfs}. 
Although adding this term together with some quadratic terms such as $(\Box\phi)^2$ and 
$\phi_{\mu\nu}\phi^{\mu\nu}$ removes the instabilities but the setup no longer describes dark 
matter at the background level. 

As we have seen, the advantage of our method is that we can control the background equations such that it
still describes a dark matter-like fluid while the model being free of instabilities at the level of 
perturbations. This can be easily achieved if we consider the simple functional form
\ba\label{f-DM}
f(\chi)=\frac{\gamma}{2}\chi^2 \, ,
\ea
for the higher derivative function in (\ref{action-Bounce-Healthy}). Imposing conditions
(\ref{Sol-beta-delta}) and (\ref{Solb9}), the background equations, given by
(\ref{Friedmann-GHD}) and (\ref{Raychuadhuri-GHD}), coincide with those studied in 
Ref. \cite{Mirzagholi:2014ifa} and therefore the model behaves like the dark matter at the
background level. At the level of perturbations, the second order actions can be obtained
by imposing condition (\ref{Solb9}) in (\ref{action2scalar-beta}) and (\ref{action2tensor-beta}) 
with $f (\chi)$ given in Eq. (\ref{f-DM}). Therefore, this model not only describes a dark matter-like
fluid at the cosmological background, but can also have healthy perturbations.

%%%%%%%%%%%%%%%%%%%%%%%%%%%%%%%%%%%%%%%%%%%%%%%%%%%%
\subsection{Nonsingular mimetic model}

In this subsection, we study nonsingular mimetic  model which is recently suggested 
 in \cite{Chamseddine:2016ktu,Chamseddine:2016uef}.
The model is defined by the action (\ref{action1}) with the following nonlinear functional 
form for the higher derivative function
\ba\label{f}
f(\chi)=\chi_m^2\left(1+\frac{1}{3}\frac{\chi^2}{\chi_m^2}-\sqrt{\frac{2}{3}}
\frac{\chi}{\chi_m} \sin^{-1}{\left(\sqrt{\frac{2}{3}}\frac{\chi}{\chi_m}\right)}
- \sqrt{1-\frac{2}{3}\frac{\chi^2}{\chi_m^2}}\,\right)\,,
\ea
where $\chi_m$ denotes the maximum value of $\chi$ such that $\chi<\sqrt{\frac{3}{2}}
\chi_m$. To ensure that general relativity arises in the low energy regime (the correspondence 
principle) we should assume that $\chi_m \sim M_{\rm Pl}$. This latter condition ensures that 
the minimum length scale determined by $l_{\rm min}=\sqrt{\frac{2}{3}} \chi_m^{-1}$ is at the 
order of Planck length $l_{\rm min}\sim l_{\rm Pl}$. 

Before explicitly confirming that the mimetic model (\ref{action1}) with higher derivative 
function defined in (\ref{f}) is nonsingular, it is insightful to qualitatively consider the model in
uniform-$\phi$ slicing. The trace of extrinsic curvature on these slices 
for FRW universe defines the expansion  rate $K=\nabla_\mu n^{\mu}=-3H$ which is also given by $K=\Box\phi=\chi$. On the other hand, the higher derivative function (\ref{f}) is defined such that $\chi$ has 
the maximum value $\chi<\sqrt{\frac{3}{2}}\chi_m$. Therefore, considering higher derivative function 
(\ref{f}) immediately implies an upper bound for the Hubble parameter. This result in some senses
signals the possibility of singularity resolution in the model. To be more precise, we  try to find an upper bound for the curvature. Indeed, working in uniform-$\phi$ slicing we 
can intuitively understand how an upper bound for  curvature would arise in this setup. In 
order to do this, we implement the Gauss-Codazzi formula $\int \sqrt{-g} R=\int \sqrt{-g} 
({^3R}+K^\mu_\nu K^\nu_\mu-K^2)$ in which $^3R$ is the spatial Ricci scalar and the boundary 
term is integrated out. We note that $^3R=0$ for the flat FRW universe and we then have 
$\int \sqrt{-g} (K^\mu_\nu K^\nu_\mu-K^2)=-\int \sqrt{-g}\,R_{\mu\nu}n^\mu n^\nu$ 
\cite{Mirzagholi:2014ifa}.  The last expression for the integral of curvature shows that the 
curvature is bounded. This is because  the trace of the extrinsic curvature is related to the second derivative 
of the scalar field via $K=\Box\phi$ and the higher derivative function (\ref{f}) is defined such 
that $\Box\phi$ has an upper bound $\Box\phi=\chi<\sqrt{\frac{3}{2}}\chi_m$. Having this in mind, 
let us calculate the background equations associated with the action (\ref{action1}). They are 
indeed the same as those in our quintic model (\ref{action-Bounce-Healthy}) when we impose 
conditions (\ref{Sol-beta-delta}) and (\ref{Solb9}). Therefore, it is sufficient to substitute (\ref{f}) 
in Eqs. (\ref{Friedmann-GHD}) and (\ref{Raychuadhuri-GHD}). 

The associated Klein-Gordon equation can be obtained by varying the action with respect to 
the scalar field $\phi$. It is not independent of Eqs. (\ref{Friedmann-GHD}) and
(\ref{Raychuadhuri-GHD}) and is given by \cite{Chamseddine:2016uef}
\ba\label{KGE-Bounce}
\frac{d}{dt}\left( 2a^3 {\bar\lambda} + 3a^3 \dot{H} f_{\chi\chi} \right)=0\,,
\ea
in which we have used the mimetic constraint (\ref{mimetic-C}). Substituting Eq. (\ref{f}) in
Eqs. (\ref{Friedmann-GHD}), (\ref{Raychuadhuri-GHD}), and (\ref{KGE-Bounce}), one can obtain the 
explicit forms of these equations. 

In order to see how these equations can support the existence of a bounce, following 
\cite{Chamseddine:2016uef}, we integrate the Klein-Gordon equation (\ref{KGE-Bounce})
and then solve for ${\bar\lambda}$ which gives 
\ba\label{lambda-KGE}
{\bar\lambda}=-\frac{c}{2a^3}-\frac{3}{2} \dot{H} f_{\chi\chi}\,,
\ea
where $c$ is a constant of integration. Substituting the above solution into the Friedmann
equation (\ref{Friedmann-GHD}) and using Eq. (\ref{f}), it is straightforward to obtain the following 
equation
\ba\label{Friedmann-LQC}
H^2=\frac{1}{3} \, \rho \, \Big(\, 1-\frac{\rho}{\rho_c}\,\Big)\,,
\ea
in which $\rho=ca^{-3}$ is the energy density and $\rho_c=2\chi_m^2$ is the maximum energy 
scale. Very interestingly, the above equation has the same form as the modified Friedmann 
equation which is suggested in loop quantum cosmology \cite{LQC1,LQC2,LQC3}. However, the modified 
Friedmann equation in loop quantum cosmology is obtained in the semiclassical 
regime of the fully quantized loop theory while there is no covariant action for them in loop 
quantum cosmology scenario. The above  coincidence therefore suggests that we can have  an effective 
higher derivative Lagrangian for the loop quantum cosmology scenarios in the context of 
mimetic gravity \cite{Bodendorfer:2017bjt,Liu:2017puc}. 

Since the background equations for the model (\ref{action1}) is the same as in our quintic model,
all of the above background analysis are also true for our quintic model 
(\ref{action-Bounce-Healthy}) if we impose conditions (\ref{Sol-beta-delta}) and (\ref{Solb9}).
Therefore, we could construct a  bouncing mimetic model by substituting (\ref{f}) in 
(\ref{action-Bounce-Healthy}). In comparison with the original nonsingular mimetic model
(\ref{action1}) which was suffering from some instabilities \cite{Firouzjahi:2017txv}, the quintic bouncing model can have healthy perturbations.

%%%%%%%%%%%%%%%%%%%%%%%%%%%%%%%%%%%%%%%%%%%%%%%%%%%%
\subsection{Randall-Sundrum cosmology via mimetic scenario}

Another interesting feature of the mimetic model (\ref{action1}) is that it can produce
background equations of the Randall-Sundrum cosmology. As a side remark, in this subsection we comment 
on the possible connection between the Randall-Sundrum cosmology and the mimetic scenario.

In Randall-Sundrum model \cite{Randall:1999ee} of brane world scenario, the Friedmann 
equation is modified  when one projects the bulk effects on the corresponding brane \cite{Cline:1999ts}. The high energy effects that arise through this projection are proportional 
to the square of the energy momentum tensor. For the case of perfect fluid which we are 
interested in cosmology, this correction will produce a term proportional to the square of the 
energy density in the associated Friedmann equations. On the other hand, we have seen that 
such a term also arises in bouncing mimetic model (\ref{Friedmann-LQC}). One then is 
persuaded to explore a possible connection between these two different scenarios. 

In the single brane version of  Randall-Sundrum  scenario the corresponding Friedmann 
equation takes the following form \cite{Maartens:2010ar}
\ba\label{Friedmann-RS}
H^2=\frac{1}{3} \rho \left(1+\frac{\rho}{2\mu}\right)\,,
\ea
where $\mu$ is the tension of the brane. Comparing the above equation with the modified 
Friedmann equation (\ref{Friedmann-LQC}) in mimetic bouncing scenario shows that 
these equations have the same form but the sign of the term $\rho^2$ is different in two models. 
We then immediately find that the Friedmann equation (\ref{Friedmann-RS}) can be obtained
from the action (\ref{action1}) if we consider the following functional form for the 
higher derivative function
\ba\label{f-RS}
f(\chi)=-\mu\left(1-\frac{1}{3}\frac{\chi^2}{\mu}+\sqrt{\frac{2}{3}}
\frac{\chi}{\sqrt{\mu}} \sinh^{-1}{\left(\sqrt{\frac{2}{3}}\frac{\chi}{\sqrt{\mu}}\right)}
- \sqrt{1+\frac{2}{3}\frac{\chi^2}{\mu}}\,\right)\,.
\ea
The above function can be obtained from its counterpart in bouncing model (\ref{f}) 
by applying the complex transformation $\chi\rightarrow\,i\chi$ and then replacing $\chi_m$ 
with $\sqrt{\mu}$. 

 Let us elaborate more on the different signs of the $\rho^2$ term in these two 
scenarios. In the nonsingular mimetic scenario the minus sign of the $\rho^2$ term in (\ref{f}) 
is crucial to get a bounce. Specifically, the minus sign in the last term of (\ref{f}) implies a 
maximum value for $\chi$ and therefore for $H$ in (\ref{Friedmann-LQC}) allowing to obtain 
a nonsingular bouncing cosmology. On the other hand, the positive sign of the $\rho^2$ term
in Randall-Sundrum model Eq. (\ref{f-RS}) shows that $\chi=\Box\phi$ can take arbitrary value 
and the Hubble parameter is not bounded. Indeed, the positive sign of $\rho^2$ in this 
scenario comes from the positive tension of the brane  $\mu>0$. The positive tension of the 
brane is equivalent to the negative curvature (negative cosmological constant) for the 
$\mbox{AdS}_5$ bulk space which prevents gravity from leaking into the extra fifth 
dimension at low energy regime \cite{Maartens:2010ar}. Therefore, apart from their 
superficial similarity, these two scenarios are very different from the cosmological point of 
view. For instance, the model (\ref{f}) leads to the nonsingular bounce while the big bang 
singularity is still present in model (\ref{f-RS}). As we have discussed in the previous 
subsection, the curvature is given by $\int \sqrt{-g} R=-\int \sqrt{-g}\,R_{\mu\nu}n^\mu n^\nu
=\int \sqrt{-g} (K^\mu_\nu K^\nu_\mu-K^2)$. In uniform-$\phi$ slicing, the trace of the 
extrinsic curvature is related to the second derivative of the scalar field as $K=\Box\phi$. In
comparison with the nonsingular model defined by the higher derivative function (\ref{f}) in
which $\Box\phi$ is bounded, there is no upper bound for $\Box\phi$ in 
Randall-Sundrum model with higher derivative function (\ref{f-RS}). Therefore, the singularity
cannot be removed in this model.

Similar to the previous subsection, all of the above background analysis are true for the quintic
model (\ref{action-Bounce-Healthy}) when we impose conditions (\ref{Sol-beta-delta}) and
(\ref{Solb9}). Therefore, we can construct a healthy counterpart of the model defined
by (\ref{f-RS}) in the context of our quintic model. 

%%%%%%%%%%%%%%%%%%%%%%%%%%%%%%%%%%%%%%%%%%%%%%%%%%%%
\section{Summary and Discussions}

In  mimetic gravity, the longitudinal mode of gravity, which is encoded in scalar 
field $\phi$, becomes dynamical which mimics the role of dark matter. The sound speed of the corresponding scalar 
cosmological perturbations however vanishes. Therefore the higher derivative term $(\Box\phi)^2$
is added to the setup which can generate a nonzero sound speed for the corresponding
cosmological perturbations. The resultant model suffers from 
some linear instabilities \cite{Ramazanov:2016xhp,Ijjas:2016pad} and it is shown that these
instabilities can be avoided by adding a coupling between the higher derivative term $\Box\phi$ 
and curvature such as $f(\Box\phi) R$ and $\nabla^\mu\nabla^\nu \phi R_{\mu\nu}$ to the setup
\cite{Hirano:2017zox,Zheng:2017qfs}. Recently, a new version of mimetic gravity has been
suggested in which the higher derivative term $(\Box\phi)^2$ is replaced with a general 
nonlinear function $f(\Box\phi)$ \cite{Chamseddine:2016uef,Chamseddine:2016ktu}. It is 
shown that the big bang singularity can be resolved in this scenario such that the 
singularity is replaced by a bounce. Interestingly, the associated cosmological equations 
coincide with those suggested by loop quantum cosmology 
\cite{Bodendorfer:2017bjt,Liu:2017puc} and therefore this mimetic model suggests an effective 
higher order action for the loop quantum cosmology scenario. Apart from these interesting 
features, similar to the mimetic dark matter model \cite{Chamseddine:2014vna}, the nonsingular 
mimetic model is also unstable \cite{Firouzjahi:2017txv}. Adding a coupling between 
the second derivative of the scalar field and curvature can make both of the mimetic dark matter 
and mimetic nonsingular model  stable. However, the background equations are modified drastically  in the presence of these terms so  whether or not they still describe dark matter or nonsingular bouncing scenarios are unclear.  

In this paper, we have worked  with classified higher derivative terms which allow us not only to make the perturbations stable but  also to control the background equations. In the first part, we have constructed the mimetic
higher order gravity model (\ref{action}) by means of curvature independent quadratic  higher
derivative terms (\ref{L2i}) and cubic curvature dependent terms (\ref{L3i}). We have shown 
that the model provides healthy scalar and tensor perturbations. In the second part, we 
have explored for the healthy extension of the original mimetic gravity (\ref{action0})  and the nonsingular mimetic  model  in our setup. We have shown that we need to consider quintic curvature dependent terms to control
both the background and perturbed equations, leading to  the quintic mimetic
model (\ref{action-Bounce-Healthy}). By appropriately choosing the coefficients of the quintic terms, we 
have shown that it is possible to keep the background equations unchanged while the cosmological perturbations may be healthy.  We then study healthy extension of mimetic dark matter and mimetic nonsingular model within our 
setup. Furthermore, we have found a possible connection between the mimetic setup and the 
Randall-Sundrum cosmology.\\

%%%%%%%%%%%%%%%%%%%%%%%%%%%%%%%%%%%%%%%%%%%%%%%%%%%%%

{\bf Acknowledgments:}  
We thank Karim Noui for useful discussions and correspondences. H. F. would like to thank Yukawa Institute for Theoretical Physics (YITP) for hospitalities while this work was in progress. 
The calculations in this work have been performed with the xAct package of Mathematica
\cite{Brizuela:2008ra,Nutma:2013zea}. M. A. Gorji thanks Ghadir Jafari for his kind assistances
with the xAct package. 
\vspace{0.7cm}

\appendix

\section{Some calculations and formulas}
\label{appA}

Some of the equations are too heavy to show in the main body of the paper and we therefore 
move them into this appendix. 

\subsection{Details of calculation of the second order action for model (\ref{action})}\label{appA-sec1}

In this section, we present some details of calculation of the second order action for the model 
(\ref{action}) in uniform-$\phi$ gauge defined by (\ref{deltaphi}), (\ref{ADM-scalar-pert-G}) and 
(\ref{ADM-scalar-pert-M}). 

Substituting from (\ref{ADM-scalar-pert-G}) and (\ref{ADM-scalar-pert-M}) in the 
ADM-decomposed action (\ref{actionADM}) and also substituting the potential from (\ref{V0}), 
$\bar{\lambda}$ from (\ref{lambda0}), and $N_1$ from (\ref{alpha}), it is straightforward to 
show that the second order action of perturbations takes the following form
\ba\label{S2}
S^{(2)} = \int d^{4}x\, a^3 \Big(L_{EH}^{(2)} + { L}_M^{(2)} + 3 \calR\, 
{L}_M^{(1)} +\frac{9}{2}\calR^2\, { L}_M^{(0)} \Big) \, ,
\ea
in which $L_{EH}^{(2)}$ represents the contribution of the Einstein Hilbert term that is given 
by 
\ba
L_{EH}^{(2)}
& = & - 3\dot{\calR}^2 - 18H \calR\dot{\calR} - \dfrac{27}{2}H^2\calR^2
- a^{-2}\big((\partial\calR)^2+2\calR\partial^2\calR\big)
\\ \nonumber 
& + & 2\dot{\calR}\partial^2\psi + 6H\left(\calR\partial^2\psi+
\partial_i\calR \partial_i\psi\right) + \partial_i\partial_j\psi\partial_i\partial_j\psi
-\left(\partial^2\psi\right)^2\, .
\ea
The rest are the contributions of the mimetic matter sector which are grouped in three 
different terms as
\ba\label{LM0-com}
L_{M}^{(0)} = - (2\dot{H} + 3 H^2) + (\alpha_1+3 \alpha_2) (6H^2+\dot{H}) 
- 6 (\kappa_1+3\kappa_2) (H \dot{H}+3 H^3) \,,
\ea
\ba\label{LM1-com}
L_{M}^{(1)} & = &  - 2(\alpha_1+3 \alpha_2) \big(-3\dot{\calR}+\partial^2\psi\big) 
+ 9 (\kappa_1+4 \kappa_2-\kappa_4) H^2 \big(-3\dot{\calR}+\partial^2\psi\big)
\\ \nonumber
& + & (\kappa_1+6 \kappa_2-3 \kappa_4) \frac{d}{dt} \big(-3\dot{\calR}+\partial^2\psi\big)
+ \frac{4}{a^2} (\kappa_1+3 \kappa_2) H \partial^2\calR \, ,
\ea
and
\ba\label{LM2-com}
L_{M}^{(2)}  &=& - (\alpha_1+3\alpha_2)\left(6H\partial_i\calR\partial_i\psi+2\dot{\calR}
\partial^2\psi-3\dot{\calR}^2\right) + \alpha_1 \partial_i\partial_j\psi\partial_i\partial_j\psi 
+ \alpha_2 (\partial^2\psi)^2 
\nonumber \\
& + & 2(\kappa_1+3\kappa_2) a^{-2} \left(\partial_i\calR\partial_i\calR+2(\dot{\calR}
-2\calR) \partial^2\calR\right) - \kappa_1 \partial_i\partial_j\psi\partial_i\partial_j
(a^{-2}\calR+\dot{\psi})
\nonumber \\
& + & 9 (\kappa_1+4 \kappa_2-\kappa_4) H \Big( 3 H \partial_i\calR\partial_i\psi 
+ (-3 \dot{\calR}+ 2 \partial^2\psi) \dot{\calR} \Big) - (\kappa_1+\kappa_2) 
a^{-2} \partial^2\calR \partial^2\psi
\nonumber \\
& + & (\kappa_1+6\kappa_2-3\kappa_4)  \Big(3\dot{H}a^{-2}\partial_i\calR
\partial_i\psi + H (6 \partial_i\dot{\calR}\partial_i\psi+3\partial_i\calR\partial_i
\dot{\psi}-\partial_i\psi\partial_i\partial^2\psi) \Big)
\nonumber \\
& - & \frac{1}{2}(\kappa_1+6\kappa_2-3\kappa_4)\frac{d}{dt}\left((-3\dot{\calR}
+2\partial^2\psi)\dot{\calR}\right) - (2\kappa_2-\kappa_4) \partial^2\psi \partial^2\dot{\psi}
\nonumber \\
& - & (2\kappa_1+11\kappa_2-2\kappa_4) H (\partial^2\psi)^2
- 3(\kappa_1+\kappa_2-\kappa_4) H \partial_i\partial_j\psi\partial_i\partial_j\psi \, .
\ea

Going to Fourier space and doing some appropriate integration bay parts, it is straightforward
to obtain the quadratic action which takes the simple form of (\ref{Lagrangian0}).

\subsection{Coefficients in equations (\ref{Friedmann-f}), (\ref{Raychuadhuri-f}), (\ref{action2scalar-beta}), (\ref{Thethat-beta}), and (\ref{cs2-beta})}\label{appA-sec2}
In this section we represent the messy expressions for the coefficients $B_i$s, $\Delta_i$s,
and $B'_i$s which are defined in equations (\ref{Friedmann-f}), (\ref{Raychuadhuri-f}), (\ref{action2scalar-beta}), (\ref{Thethat-beta}), and (\ref{cs2-beta}). 

The coefficients $B_i$s and $\Delta_i$s in equations (\ref{Friedmann-f}) and 
(\ref{Raychuadhuri-f}) are given by
\ba\label{Bi}
&&B_1 = 2 \beta_1+12 \beta_2-6 \beta_4-\beta_6+3 \beta_7+4 \beta_9 +4 \beta_{11} \,, \\
&&B_2 = 9 \beta_1 + 24 \beta_2+ 3 \beta_4 - 3 \beta_7 + 6 \beta_9 + 4 \beta_{11} \,,
\nonumber \\
&&B_3 = 27 \beta_1 + 12 \beta_2+ 69 \beta_4 + 9 \beta_6 - 42 \beta_7- 6 \beta_9 
- 16 \beta_{11} \,,
\nonumber \\
&&B_4 = 18 \beta_1+48 (\beta_2+ \beta_3)-18 \beta_4-12\beta_5-6 \beta_6
+6 \beta_7+3 \beta_8 \nonumber \\
&&\hspace{.6cm}+12(\beta_9+ \beta_{10}) +8 (\beta_{11}+\beta_{12}) \,,
\nonumber \\
&&B_5 = 30 \beta_1 + 3 (40 \beta_2-16 \beta_3-18 \beta_4+6 \beta_5-4 
\beta_6+8 \beta_7-2\beta_8+12 \beta_9) \nonumber \\
&&\hspace{.6cm} -4( 3\beta_{10} - 8\beta_{11} + 2 \beta_{12}) \,,
\nonumber \\
&&B_6 = 48\beta_1 + 3 (32 \beta_2-44 \beta_3+18 \beta_5-4 \beta_6-8\beta_7
-7 \beta_8+8 \beta_9) \nonumber \\
&&\hspace{.6cm} - 4(9 \beta_{10}-4 \beta_{11} +7 \beta_{12}) \,,
\nonumber \\
&&B_7 = 54 \beta_1+ 204 \beta_2- 168 \beta_3- 24 \beta_4+ 51 \beta_5
- 6 \beta_6 - 3 \beta_7- 15 \beta_8 \nonumber \\
&&\hspace{.6cm}+ 60 \beta_9 - 42 \beta_{10}+ 52 \beta_{11}- 28 \beta_{12} \,,
\nonumber \\
&&B_8 = 6 (\beta_1 + 2 \beta_2) - 3 (4 \beta_3+2\beta_4-2 \beta_5+\beta_6
-\beta_7+\beta_8) \nonumber \\
&&\hspace{.6cm}+4 (\beta_9-\beta_{10}+ \beta_{11}- \beta_{12})\,, \nonumber
\ea
and 
\ba\label{deltai}
&&\Delta_1 = 9 \delta_1+3 \delta_2+\delta_3-90 \delta_{11}-30 \delta_{12}-10 
\delta_{13}-8 \delta_{21} \,, \\
&&\Delta_2 = -27 \delta_1 - 15 \delta_2 -3 (2 \delta_3+3 \delta_4+\delta_6) 
+90 \delta_{11}+54 \delta_{12}+22\delta_{13}
\nonumber \\
&&\hspace{.6cm}
+36 \delta_{14}+12\delta_{16}+6\delta_{21} +2 \delta_{22} \,,
\nonumber \\
&&\Delta_3 = 9 \delta_1+\delta_2-3 \delta_4-\delta_6-18 \delta_{11}-2\delta_{12}
+6 \delta_{14}+2 \delta_{16}-2\delta_{21} \,,
\nonumber \\
&&\Delta_5 = 54 \delta_1-24 \delta_2-3 (5 \delta_3+21 \delta_4+7 \delta_6)- 
432 \delta_{11}-12 \delta_{12}+18 \delta_{13}\nonumber \\
&&\hspace{.6cm}+198 \delta_{14}+66 \delta_{16}-50 \delta_{21}+8 \delta_{22} \,, \nonumber 
\ea

while the coefficients $B'_i$s in equations (\ref{action2scalar-beta}), (\ref{Thethat-beta}), 
and (\ref{cs2-beta}) are given by

\begin{eqnarray}\label{Bprimes}
&&B'_1=2{{\beta }_{1}}-{{\beta }_{6}} \, , \hspace{5cm}
B'_{2}=3{{\beta }_{1}}+12{{\beta }_{2}}-3{{\beta }_{4}}
-{{\beta }_{6}}+3{{\beta }_{9}} \, , \\
&&B'_3=4{{\beta }_{1}}-{{\beta }_{6}}+3\left( 2{{\beta }_{9}}
+8{{\beta }_{2}} \right) \, , \hspace{2cm} 
B'_4=4{{\beta }_{1}}+3\left( 6{{\beta }_{4}}+{{\beta }_{6}}
-2{{\beta }_{9}}-8{{\beta }_{2}} \right) \, , \nonumber \\
&&B'_{5}=3{{\beta }_{4}}+{{\beta }_{6}}+9{{\delta }_{11}}+3{{\delta }_{12}}
+{{\delta }_{13}} \, , \hspace{1.5cm}
B'_6=-2{{\beta }_{1}}+{{\beta }_{6}}+2{{\beta }_{9}} \, , \nonumber\\
&&B'_7=14{{\beta }_{1}}+48{{\beta }_{2}}+12{{\beta }_{4}}+3{{\beta }_{6}}
+2{{\beta }_{9}}+72{{\delta }_{11}}+24{{\delta }_{12}}+8{{\delta }_{13}} \, , \nonumber\\
&&B'_{8}=4{{\beta }_{1}}+24{{\beta }_{2}}+6{{\beta }_{4}}+3{{\beta }_{6}}
+41{{\beta }_{9}}+9{{\delta }_{11}}+3{{\delta }_{12}}+{{\delta }_{13}} \, , \nonumber\\
&&B'_{9}=49{{\beta }_{1}}+108{{\beta }_{2}}+27{{\beta }_{4}}+3{{\beta }_{6}}
+20{{\beta }_{9}}+27{{\delta }_{1}}+6\left( {{\delta }_{12}}+{{\delta }_{13}}
+{{\delta }_{2}} \right) \, , \nonumber\\
&&B'_{10}=205{{\beta }_{1}}+528{{\beta }_{2}}+114{{\beta }_{4}}+9{{\beta }_{6}}
+92{{\beta }_{6}}+54{{\delta }_{1}}+126{{\delta }_{11}}+66{{\delta }_{12}}
+38{{\delta }_{13}}+12{{\delta }_{2}} \, , \nonumber\\
&&B'_{11}=103{{\beta }_{1}}-204{{\beta }_{2}}+33{{\beta }_{4}}+32{{\beta }_{9}}
-27{{\delta }_{1}}+126{{\delta }_{11}}+48{{\delta }_{12}}+20{{\delta }_{13}}
-6{{\delta }_{2}} \, , \nonumber \\
&&B'_{12}=107{{\beta }_{1}}+84{{\beta }_{2}}+141{{\beta }_{4}}+19{{\beta }_{5}}
+9\left( 9{{\delta }_{1}}+2\left( {{\delta }_{12}}+{{\delta }_{13}}+{{\delta }_{2}} 
\right) \right)\nonumber\\
&&B'_{13}=19{{\beta }_{1}}+60{{\beta }_{2}}+57{{\beta }_{4}}+8{{\beta }_{6}}
+81{{\delta }_{1}}-54{{\delta }_{11}}-36{{\delta }_{12}}-24{{\delta }_{13}}
+18{{\delta }_{2}}\nonumber\\
&&B'_{14}=302{{\beta }_{1}}+192{{\beta }_{2}}+366{{\beta }_{4}}+49{{\beta }_{6}}
+162{{\delta }_{1}}+54{{\delta }_{11}}+90{{\delta }_{12}}+78{{\delta }_{13}}
+36{{\delta }_{2}} \nonumber\\
&&B'_{15}=4{{\beta }_{1}}+6{{\beta }_{4}}+{{\beta }_{6}} \nonumber\\
&&B'_{16}=76{{\beta }_{1}}+240{{\beta }_{2}}+21{{\beta }_{4}}-{{\beta }_{6}}
+27\left( 9{{\delta }_{11}}+3{{\delta }_{12}}+{{\delta }_{13}} \right)\nonumber\\
&&B'_{17}=4{{\beta }_{1}}+24{{\beta }_{2}}+21{{\beta }_{4}}+8{{\beta }_{6}}
+9\left( 9{{\delta }_{11}}+3{{\delta }_{12}}+{{\delta }_{13}} \right)\nonumber\\
&&B'_{18}=28{{\beta }_{1}}+96{{\beta }_{2}}+21{{\beta }_{4}}+5\left( {{\beta }_{6}}
+27{{\delta }_{11}}+9{{\delta }_{12}}+3{{\delta }_{13}} \right)\nonumber\\
&&B'_{19}=5{{\beta }_{1}}+12{{\beta }_{2}}-3{{\beta }_{4}}-2{{\beta }_{6}}\nonumber \, .
\end{eqnarray}

%%%%%%%%%%%%%%%%%%%%%%%%%%%%%%%%%%%%%%%%%%%%%%%%%%
\section{Cubic and quartic mimetic models with unchanged background equations}

In this appendix, we show that the instabilities in nonsingular mimetic model (\ref{action1})
cannot be completely removed by adding the cubic and quartic curvature dependent terms if we 
demand that the background equations continue to take the simple forms (\ref{Friedmann-GHD}) and
(\ref{Raychuadhuri-GHD}). This is the reason for which we have considered the quintic terms in 
section 3.

We add the cubic and quartic terms to the action of nonsingular mimetic model (\ref{action1}) as follows 
\ba\label{action-Bounce-L34}
S= \int d^4 x  \sqrt{-g} \left[ \frac{1}{2} R + \lambda \left(X+ 1 \right)+ f(\chi) 
+ \sum_{i=1}^5\,\kappa_i L^{(1,2)}_i + \sum_{i=1}^{13}\,\sigma_i L^{(2,2)}_i \right] \, ,
\ea
in which $\kappa_i$ label the cubic curvature dependent terms $L^{(1,2)}_i$ that are given by 
(\ref{L3i}) while $\sigma_i$ label the quartic curvature dependent terms $L^{(2,2)}_i$ that are 
defined as
\ba\label{L4i}
&&L_1^{(2,2)} = \phi_{\mu\nu} \phi^{\mu\nu} R \, , \hspace{3cm} L_2^{(2,2)} = 
(\Box\phi)^2 R \, , \hspace{2cm} L_3^{(2,2)} = (\Box\phi) \phi^\mu \phi_{\mu\nu} \phi^{\nu} 
R \, , \nonumber \\  
&&L_4^{(2,2)} = \phi^{\mu} \phi_{\mu\rho} \phi^{\rho\nu} \phi_{\nu} R \, , \hspace{2.15cm}
L_5^{(2,2)} =  (\phi^\mu \phi_{\mu\nu} \phi^{\nu})^2 R \, , \hspace{1.1cm}
L_6^{(2,2)} = \phi_{\mu\nu} \phi^{\mu\nu} \phi^\alpha \phi^\beta R_{\alpha,\beta} \, , 
\nonumber \\
&& L_7^{(2,2)} = (\Box\phi)^2 \phi^\alpha \phi^\beta R_{\alpha,\beta} \, , 
\hspace{1.8cm} L_8^{(2,2)} = (\Box\phi) \phi^\mu \phi_{\mu\nu} \phi^{\nu} 
\phi^\alpha \phi^\beta R_{\alpha,\beta} \, , \nonumber \\  
&&L_9^{(2,2)} = \phi^{\mu} \phi_{\mu\rho} \phi^{\rho\nu} \phi_{\nu} 
\phi^\alpha \phi^\beta R_{\alpha,\beta} , \hspace{.9cm}
L_{10}^{(2,2)} =  (\phi^\mu \phi_{\mu\nu} \phi^{\nu})^2 \phi^\alpha \phi^\beta 
R_{\alpha,\beta} \, , \\
&&L_{11}^{(2,2)} = (\Box\phi) R_{\alpha\beta} \phi^{\alpha\beta} \, , \hspace{2.3cm}
L_{12}^{(2,2)} = \phi^\mu \phi_{\mu\nu} \phi^{\nu} R_{\alpha\beta} \phi^{\alpha\beta} \, , 
\hspace{.7cm}
L_{13}^{(2,2)} = \phi^{\alpha\mu} \phi^{\beta\nu} R_{\mu\nu\alpha\beta} \, . \nonumber
\ea
The above are the independent quartic curvature terms which we have found so far. Some other
independent terms may still be found.  However, we show that the quartic 
terms cannot generally make the setup  (\ref{action1}) stable for arbitrary form of function 
$f(\Box\phi)$ if we demand the  background equations to remain unchanged. 

\subsection{Background equations}

Varying the action (\ref{action-Bounce-L34}) with respect to the metric, we obtain the Einstein 
equations which now have very complicated forms. In cosmological background (\ref{FRW}) they lead 
to the following modified Friedmann and Raychuadhuri equations 
\ba\label{Friedmann-L34}
3H^2 &=& -2 {\bar\lambda} -( f + 3H f_{\chi} + 3 \dot{H} f_{\chi\chi} ) - 
9 (\kappa_1+2 \kappa_2+4 \kappa_3-\kappa_4-\kappa_5) H^3 \\
&+& 3 (4\kappa_1+8\kappa_2-14 \kappa_3+4 \kappa_4+5\kappa_5) H\dot{H}
+3 (\kappa_1+2\kappa_2-2\kappa_3-\kappa_4+\kappa_5) \ddot{H} \,, \nonumber \\
&+&3 \big(42\sigma_1+6 (9 \sigma_2+6\sigma_3-2\sigma_6-3 \sigma_7)
-9 (\sigma_8 -2\sigma_{11}-\sigma_{12})+4 \sigma_{13} \big) H^4 \nonumber \\
&+&18 \big(2\sigma_1-12 \sigma_2+9 \sigma_3-2\sigma_6-3 \sigma_8 
-4\sigma_{11}+2\sigma_{12}-2\sigma_{13}\big) H^2\dot{H} \nonumber \\
&-&3 \big(3 (4\sigma_2
-2\sigma_3-2\sigma_7+\sigma_8)+ 4\sigma_{11}-\sigma_{12}+2\sigma_{13} \big) (\dot{H}^2+H\ddot{H}) \, , \nonumber
\ea
and
\ba\label{Raychuadhuri-L34}
2 \dot{H}+3H^2 &=& - (f+ 3H f_{\chi} -3 \dot{H}  f_{\chi\chi})
- 9 (\kappa_1+2\kappa_2+\kappa_4) (H\dot{H}+H^3) \nonumber \\
&+& 6 (3\sigma_1+9\sigma_2+3\sigma_{11}+\sigma_{13})(4\dot{H} H^2 + 3H^4) \, .
\ea

In order to obtain the nonsingular mimetic  cosmology with the Friedmann equation 
(\ref{Friedmann-LQC}), we should choose the coefficients $\kappa_i$ and $\theta_i$ such
that (\ref{Friedmann-L34}) and (\ref{Raychuadhuri-L34}) coincide with (\ref{Friedmann-GHD}) 
and (\ref{Raychuadhuri-GHD}) respectively. In other words, we should fix $\kappa_i$ and 
$\sigma_i$ such that the corrections labeled by $\kappa_i$ and $\sigma_i$, which are originated 
from the terms $L^{(1,2)}_i$ and $L^{(2,2)}_i$, do not contribute to the background 
equations. Note that they could still affect the linear equations at the level of perturbations 
even if we demand they do not affect the background equations.  

Looking at (\ref{Raychuadhuri-L34}) we therefore demand
\ba\label{Raychuadhuri-Conditions-L34}
\kappa_1+2\kappa_2+\kappa_4 = 0 \, , \hspace{2cm}
3\sigma_1+9\sigma_2+3\sigma_{11}+\sigma_{13} = 0 \, ,
\ea
and from (\ref{Friedmann-L34}) we get
\ba\label{Friedmann-Conditions-L34}
&& \kappa_1+2 \kappa_2+4 \kappa_3-\kappa_4-\kappa_5=0 \, , \hspace{1cm}
4\kappa_1+8\kappa_2-14 \kappa_3+4 \kappa_4+5\kappa_5=0 \, , \nonumber \\
&& \kappa_1+2\kappa_2-2\kappa_3-\kappa_4+\kappa_5=0 \, , \nonumber \\
&&42\sigma_1+6 (9\sigma_2+6\sigma_3-2\sigma_6-3 \sigma_7)
-9 \sigma_8+18\sigma_{11}+9\sigma_{12}+4 \sigma_{13}=0 \, , \nonumber \\
&& 2\sigma_1-12 \sigma_2+9 \sigma_3-2\sigma_6
-3 \sigma_8-4\sigma_{11}+2\sigma_{12}-2\sigma_{13}=0 \, , \nonumber \\
&&3 (4\sigma_2-2 \sigma_3 -2 \sigma_7+\sigma_8)
+4\sigma_{11}-\sigma_{12}+2\sigma_{13}=0 \, .
\ea

Solving the above equations for $\kappa_i$ give
\ba\label{kappaiSol-L3}
\kappa_2=-\frac{1}{2}\kappa_1 \, , \hspace{1cm} \kappa_3=\kappa_4=\kappa_5=0 \,.
\ea

Solving for $\sigma_{13}$ from (\ref{Raychuadhuri-Conditions-L34}) and then substituting the
result in (\ref{Friedmann-Conditions-L34}), we can express three of nine nonzero coefficients
$\sigma_i$ in terms of the other six ones. Without loss of generality, we solve for $\sigma_{11}$,
$\sigma_{12}$ and $\sigma_{13}$ which give
\ba\label{thetaiSol}
&& \sigma_{13}=\frac{3}{2}(2\sigma_1+3\sigma_3+2\sigma_6+12\sigma_7-3\sigma_8) \, , 
\hspace{1cm} \sigma_{12}=-2\sigma_1-3\sigma_3+2\sigma_6+6\sigma_7 \,, \nonumber \\
&& \sigma_{11}=-\frac{1}{2}(4\sigma_1+6\sigma_2+3\sigma_3+2\sigma_6+12\sigma_7
-3\sigma_8) \,.
\ea
Thus, to have a bouncing cosmology with background equations (\ref{Friedmann-GHD}) and
(\ref{Raychuadhuri-GHD}) in model (\ref{action-Bounce-L34}), there are six free parameters
$\sigma_1$, $\sigma_2$, $\sigma_3$, $\sigma_6$, $\sigma_7$, and $\sigma_8$ associated to
the quartic terms (\ref{L4i}) while there are only one free parameter $\kappa_1$ for the case 
of cubic terms (\ref{L3i}). This result show that fixing the background equations to have
nonsingular mimetic cosmology with Eq. (\ref{Friedmann-LQC}), makes the space of parameters
very restricted for the case of cubic terms. This is indeed the reason for which the cubic 
terms cannot make the setup stable. For the case of quartic terms, however, the space of
parameters is larger and we then hope to find a stable solution for the mimetic nonsingular
model. But, as we will show in the next subsection, these terms also cannot help either.

\subsection{Cosmological perturbations and stability analysis: Cubic terms}

Let us first consider cubic terms and turn off quartic terms $\sigma_i=0$ for a while. It is 
then straightforward to obtain the second order action for the scalar perturbations for the
model (\ref{action-Bounce-L34}) which is given by

\ba\label{action2scalar-L3}
S^{(2)} = \frac{1}{2} \int d^3 k \int dt\,
a^3 \left[ (3 f_{\chi\chi}-2) (3\dot{\calR}_k+ 2 k^2 \psi_k ) \dot{\calR}_k
+2a^{-2}k^2\calR_k^2 + (f_{\chi\chi}+8\kappa_1 H) k^4 \psi_k^2 \right] ,
\ea
in which we have substituted (\ref{kappaiSol-L3}). Moreover, we have used the background
equations (\ref{Friedmann-GHD}) and (\ref{Raychuadhuri-GHD}) in order to remove
$\bar{\lambda}$ and $\dot{H}$. To be more precise, we must use background equations 
(\ref{Friedmann-L34}) and (\ref{Raychuadhuri-L34}). But, these equations reduce to 
(\ref{Friedmann-GHD}) and (\ref{Raychuadhuri-GHD}) respectively when $\sigma_i=0$ and 
we impose (\ref{kappaiSol-L3}). 

In action (\ref{action2scalar-L3}), $\psi_k$ appears with no time derivative and so we can
eliminate it by means of its equation of motion
\ba\label{psiSol-L3}
\psi_k=  - \left(\frac{3f_{\chi\chi}-2}{f_{\chi\chi}+8\kappa_1 H} \right) k^{-2}\dot{\calR}_k\,.
\ea
The above relation shows that $\psi_k \sim \dot{\calR}_k$ and therefore it does not contribute 
to the gradient term $\calR_k^2$ after substituting the above solution into the action
(\ref{action2scalar-L3}). Note that for the model (\ref{action}) in section 2, the solution for
$\psi_k$ given by (\ref{psi}) was proportional to both $\dot{\calR}_k$ and ${\calR}_k$ 
and therefore it affected both the kinetic term and the gradient term. To be more precise, the
coefficient of ${\calR}_k$ in (\ref{psi}) was $(\kappa_1+2\kappa_2)$ which we demand to be
zero in (\ref{kappaiSol-L3}) in order to keep the background equations unchanged. This is 
the reason why the cubic term cannot make the model (\ref{action1}) stable assuming 
that the background equations remain unchanged. 

Substituting $\psi_k$ into the action
(\ref{action2scalar-L3}) we obtain the reduced action in term of curvature perturbation as follows
\ba\label{S2-L3}
S^{(2)} = \int d^3 k \int dt\,
a^3 \left[ (1+12\kappa_1H) \left(\frac{3 f_{\chi\chi}-2}{f_{\chi\chi}+8\kappa_1H}\right)
\dot{\calR}_k^2+\frac{k^2}{a^2}\,\calR_k^2 \right] \, .
\ea
In the above action the gradient term appears with the wrong sign. Indeed, it was not affected 
by the cubic curvature dependent terms and therefore there is always linear instability for the
model (\ref{action-Bounce-L34}) if we consider only cubic terms (\ref{L3i}). The kinetic part in 
(\ref{S2-L3}) is however affected by these terms such that the ghost instability can be avoided.

\subsection{Cosmological perturbations and stability analysis: Quartic terms}

In the previous subsection, we have shown that the cubic terms could remove the ghost instability 
in (\ref{action-Bounce-L34}) but it cannot cure the  gradient term if we demand the background 
equations to remain unchanged. Thus, in this subsection, we neglect them ($\kappa_i=0$) and 
focus on the quartic terms (\ref{L4i}). 

Following the same steps as in the previous subsection, it is straightforward to show that the 
second order action for the scalar perturbations in Fourier space will be

\ba\label{action2scalar-theta}
S^{(2)} &=& \frac{1}{2} \int d^3 k \int dt\,
a^3 \bigg[(3 f_{\chi\chi}-2) (3\dot{\calR}_k+ 2 k^2 \psi_k ) \dot{\calR}_k
+2 a^{-2} k^2 \calR_k^2 + f_{\chi\chi} k^4 \psi_k^2 
\nonumber \\
&-& 2\Sigma_1 a^{-2}H\, k^4\psi_k \calR_k 
+(\Sigma_2+\Sigma_3+\Sigma_4) \Big(\frac{f+ 3 H f_{\chi}
+3 H^2 f_{\chi\chi}+H^2}{3 f_{\chi\chi}-2}\Big) k^4\psi_k^2\,\bigg] \,, 
\ea
in which we have used the background equations (\ref{Friedmann-GHD}) and
(\ref{Raychuadhuri-GHD}) in order to eliminate ${\bar{\lambda}}$ and $\dot{H}$ and also 
we have defined 
\ba\label{Thetai}
&&\Sigma_1=6 \sigma_2-3 \sigma_3-2\sigma_6-12 \sigma_7+3 \sigma_8 \, , \nonumber \\
&&\Sigma_2= 8 \sigma_1+30 \sigma_2-27\sigma_3-38 \sigma_6-156 \sigma_7+27 \sigma_8 \, ,
\nonumber \\
&&\Sigma_3= 12\sigma_1+6 \sigma_2+3\sigma_3-2 \sigma_6+12 \sigma_7-3\sigma_8 \, ,
\nonumber \\
&&\Sigma_4= 20 \sigma_1-42 \sigma_2+63 \sigma_3+70 \sigma_6+348 \sigma_7-63 \sigma_8 \, ,
\ea
In action (\ref{action2scalar-theta}), $\psi_k$ appears with no time derivative and it can 
be removed by its equation of motion
\ba\label{psi-L4}
\psi_k = \frac{(3 f_{\chi\chi}-2)\,\big(\Sigma_1 a^{-2}H\,\calR_k
- (3 f_{\chi\chi}-2) k^{-2} \dot{\calR}_k \big)}{ (3 f_{\chi\chi}-2) f_{\chi\chi} 
+(\Sigma_2+\Sigma_3+\Sigma_4) (f+ 3 H f_{\chi}+3 H^2f_{\chi\chi} +H^2)} \, .
\ea

Substituting the above solution into the action (\ref{action2scalar-theta}), we obtain the
reduced action in term of curvature perturbation. Before doing this, let us elaborate more
on solution (\ref{psi-L4}) and compare it with its counterpart (\ref{psiSol-L3}) where 
we only considered the cubic terms. In (\ref{psi-L4}), $\psi_k$ is proportional to both the
curvature perturbation $\calR_k$ and its time derivative $\dot{\calR}_k$ while it was only 
proportional to $\dot{\calR}_k$ in (\ref{psiSol-L3}). Thus, substituting (\ref{psi-L4}) into the
action (\ref{action2scalar-theta}) changes both the kinetic and gradient terms while
(\ref{psiSol-L3}) cannot affect the gradient term after the substitution in 
(\ref{action2scalar-L3}). This is the reason for which the gradient term in (\ref{S2-L3}) was
exactly the same as in standard mimetic scenarios. On the other hand, looking at the numerator 
of solution (\ref{psi-L4}), we can see that the quartic terms does not contribute to the term
proportional to $\dot{\calR}_k$. Remember the model (\ref{action}) in which in the  solution for 
$\psi_k$, given by (\ref{psi}), there was a separate contribution for $\dot{\calR}_k$ from 
quadratic terms label by $\alpha_i$. In this respect, the kinetic term gets contribution only 
from the $\psi_k^2$ term in (\ref{action2scalar-theta}) while the gradient term gets 
contribution not only from $\psi_k^2$ term but also from $\calR_k \psi_k$ term in 
(\ref{action2scalar-theta}). Therefore, we cannot separately control both the kinetic and
gradient terms. However, the setup may provide healthy perturbations for some functional
forms of $f(\chi)$. For instance, our analysis show that the setup can provide healthy
perturbations for the case of $f(\chi)\sim \chi^2$ while it is sick for some values of wave
number $k$ in the bouncing mimetic model with nonlinear higher
derivative function (\ref{f}). In summary, we cannot conclude that quartic model 
(\ref{action-Bounce-L34}) with background equations (\ref{Friedmann-GHD}) and 
(\ref{Raychuadhuri-GHD}) is stable for general form of higher derivative function 
$f(\chi)$.

Let us now look at the tensor modes. It is straightforward to show that the second order
action of the tensor perturbations for the model (\ref{action-Bounce-L34}) in the absence of the
cubic terms $\kappa_i=0$ is given by
\ba\label{action2tensor-theta}
S^{(2)} = \frac{1}{4} \sum_{s=+,\times} \int d^3 k \int dt \,
\Theta_t a^3 \Big(\dot{h}^s_k\dot{h}^s_k-a^{-2}k^2c_t^2h^s_kh^s_k\Big) \, ,
\ea
where we have defined 
\ba\label{Thethat-theta}
\Theta_t= 1 + \frac{\Sigma_3 (3f+f_{\chi}) + 9 \Sigma_5 H^2 f_{\chi\chi}
+ 3 \Sigma_6 H^2}{2 (3 f_{\chi\chi}-2)} \, ,
\ea
with 
\ba\label{Thetai-Tensor}
&&\Sigma_5 = 8 \sigma_1-6 \sigma_2-9 \sigma_3-14 \sigma_6-48 \sigma_7+9\sigma_8 \, , 
\nonumber \\
&& \Sigma_6 = 20\sigma_1+30 \sigma_2+27 \sigma_3+22 (\sigma_6+6 \sigma_7)-27 
\sigma_8 \, ,
\ea
and the speed of gravitational waves as
\ba\label{cs2-theta}
c_t^2=\Theta_t^{-1}\left[1- \frac{3}{2}\Sigma_1 
\left(\frac{f+3 H f_{\chi\chi} + 3 H^2 f_{\chi\chi}+H^2}{3 f_{\chi\chi}-2}\right) \right] \, .
\ea

Looking at both second order actions for scalar (\ref{action2scalar-theta}) and tensor
(\ref{action2tensor-theta}) perturbations, we see that there are six different 
combinations of six free parameters defined by $\Theta_i$ in (\ref{Thetai}) and 
(\ref{Thetai-Tensor}). Thus, the space of parameters is sufficiently large when we consider 
the quartic terms (\ref{L4i}) in model (\ref{action-Bounce-L34}). For the case of tensor
perturbations with the action (\ref{action2tensor-theta}), we can control both the kinetic and
gradient terms and therefore it is possible to have healthy tensor modes.

Therefore, instead of the model (\ref{action-Bounce-L34}) defined by the cubic (\ref{L3i}) and
quartic (\ref{L4i}) terms, in section 3, we have introduced the  model  (\ref{action-Bounce-Healthy}) 
with the quintic terms (\ref{L5i32}) and  (\ref{L5i14}). Indeed, this model has many more parameters 
to have healthy scalar and  tensor perturbations.

%%%%%%%%%%%%%%%%%%%%%%%%%%%%%%%%%%%%%%%%%%%%%%%%%%
{}


\vspace{0.7cm}

\begin{thebibliography}{}

%\cite{Chamseddine:2013kea}
\bibitem{Chamseddine:2013kea} 
A.~H.~Chamseddine and V.~Mukhanov,
%``Mimetic Dark Matter,''
JHEP {\bf 1311}, 135 (2013),
% doi:10.1007/JHEP11(2013)135
[arXiv:1308.5410 [astro-ph.CO]].
%%CITATION = doi:10.1007/JHEP11(2013)135;%%
%89 citations counted in INSPIRE as of 21 Jan 2017

%\cite{Bekenstein:1992pj}
\bibitem{Bekenstein:1992pj} 
J.~D.~Bekenstein,
%``The Relation between physical and gravitational geometry,''
Phys.\ Rev.\ D {\bf 48}, 3641 (1993), 
%doi:10.1103/PhysRevD.48.3641
[gr-qc/9211017].
%%CITATION = doi:10.1103/PhysRevD.48.3641;%%
%189 citations counted in INSPIRE as of 01 Jul 2017

%\cite{Deruelle:2014zza}
\bibitem{Deruelle:2014zza} 
N.~Deruelle and J.~Rua,
%``Disformal Transformations, Veiled General Relativity and Mimetic Gravity,''
JCAP {\bf 1409}, 002 (2014), 
%doi:10.1088/1475-7516/2014/09/002
[arXiv:1407.0825 [gr-qc]].
%%CITATION = doi:10.1088/1475-7516/2014/09/002;%%
%53 citations counted in INSPIRE as of 01 Jul 2017

%\cite{Domenech:2015tca}
\bibitem{Domenech:2015tca} 
G.~Domènech, S.~Mukohyama, R.~Namba, A.~Naruko, R.~Saitou and Y.~Watanabe,
%``Derivative-dependent metric transformation and physical degrees of freedom,''
Phys.\ Rev.\ D {\bf 92}, no. 8, 084027 (2015), 
%doi:10.1103/PhysRevD.92.084027
[arXiv:1507.05390 [hep-th]].
%%CITATION = doi:10.1103/PhysRevD.92.084027;%%
%33 citations counted in INSPIRE as of 01 Jul 2017


%\cite{Arroja:2015wpa}
\bibitem{Arroja:2015wpa} 
F.~Arroja, N.~Bartolo, P.~Karmakar and S.~Matarrese,
%``The two faces of mimetic Horndeski gravity: disformal transformations and Lagrange multiplier,''
JCAP {\bf 1509}, 051 (2015), 
%doi:10.1088/1475-7516/2015/09/051
[arXiv:1506.08575 [gr-qc]].
%%CITATION = doi:10.1088/1475-7516/2015/09/051;%%
%30 citations counted in INSPIRE as of 08 Jul 2017

%\cite{Motohashi:2015pra}
\bibitem{Motohashi:2015pra} 
  H.~Motohashi and J.~White,
  %``Disformal invariance of curvature perturbation,''
  JCAP {\bf 1602}, no. 02, 065 (2016),
 % doi:10.1088/1475-7516/2016/02/065
  [arXiv:1504.00846 [gr-qc]].
  %%CITATION = doi:10.1088/1475-7516/2016/02/065;%%
  %19 citations counted in INSPIRE as of 18 Nov 2017

%\cite{Takahashi:2017zgr}
\bibitem{Takahashi:2017zgr} 
  K.~Takahashi, H.~Motohashi, T.~Suyama and T.~Kobayashi,
  %``General invertible transformation and physical degrees of freedom,''
  Phys.\ Rev.\ D {\bf 95}, no. 8, 084053 (2017), 
  %doi:10.1103/PhysRevD.95.084053
  [arXiv:1702.01849 [gr-qc]].
  %%CITATION = doi:10.1103/PhysRevD.95.084053;%%
  %2 citations counted in INSPIRE as of 18 Nov 2017

%\cite{Achour:2016rkg}
\bibitem{Achour:2016rkg} 
J.~Ben Achour, D.~Langlois and K.~Noui,
%``Degenerate higher order scalar-tensor theories beyond Horndeski and disformal transformations,''
Phys.\ Rev.\ D {\bf 93}, no. 12, 124005 (2016),
%doi:10.1103/PhysRevD.93.124005
[arXiv:1602.08398 [gr-qc]].
%%CITATION = doi:10.1103/PhysRevD.93.124005;%%
%32 citations counted in INSPIRE as of 01 Jul 2017

%\cite{Ramazanov:2016xhp}
\bibitem{Ramazanov:2016xhp} 
S.~Ramazanov, F.~Arroja, M.~Celoria, S.~Matarrese and L.~Pilo,
%``Living with ghosts in Hořava-Lifshitz gravity,''
JHEP {\bf 1606}, 020 (2016),
%doi:10.1007/JHEP06(2016)020
[arXiv:1601.05405 [hep-th]].
%%CITATION = doi:10.1007/JHEP06(2016)020;%%
%13 citations counted in INSPIRE as of 28 May 2017

\bibitem{MimeticNew}
%\cite{Momeni:2014qta}
%\bibitem{Momeni:2014qta} 
D.~Momeni, A.~Altaibayeva and R.~Myrzakulov,
%``New Modified Mimetic Gravity,''
Int.\ J.\ Geom.\ Meth.\ Mod.\ Phys.\  {\bf 11}, 1450091 (2014),
%doi:10.1142/S0219887814500911
[arXiv:1407.5662 [gr-qc]].
%%CITATION = doi:10.1142/S0219887814500911;%%
%33 citations counted in INSPIRE as of 08 Jul 2017

%\cite{Leon:2014yua}
%\bibitem{Leon:2014yua} 
G.~Leon and E.~N.~Saridakis,
%``Dynamical behavior in mimetic F(R) gravity,''
JCAP {\bf 1504}, no. 04, 031 (2015),
%doi:10.1088/1475-7516/2015/04/031
[arXiv:1501.00488 [gr-qc]].
%%CITATION = doi:10.1088/1475-7516/2015/04/031;%%
%35 citations counted in INSPIRE as of 08 Jul 2017

%\cite{Matsumoto:2015wja}
%\bibitem{Matsumoto:2015wja} 
J.~Matsumoto, S.~D.~Odintsov and S.~V.~Sushkov,
%``Cosmological perturbations in a mimetic matter model,''
Phys.\ Rev.\ D {\bf 91}, no. 6, 064062 (2015),
%doi:10.1103/PhysRevD.91.064062
[arXiv:1501.02149 [gr-qc]].
%%CITATION = doi:10.1103/PhysRevD.91.064062;%%
%33 citations counted in INSPIRE as of 08 Jul 2017

%\cite{Haghani:2015iva}
%\bibitem{Haghani:2015iva} 
Z.~Haghani, T.~Harko, H.~R.~Sepangi and S.~Shahidi,
%``Cosmology of a Lorentz violating Galileon theory,''
JCAP {\bf 1505}, 022 (2015),
%doi:10.1088/1475-7516/2015/05/022
[arXiv:1501.00819 [gr-qc]].
%%CITATION = doi:10.1088/1475-7516/2015/05/022;%%
%13 citations counted in INSPIRE as of 08 Jul 2017

%\cite{Momeni:2015gka}
%\bibitem{Momeni:2015gka} 
D.~Momeni, R.~Myrzakulov and E.~Güdekli,
%``Cosmological viable mimetic $f(R)$ and $f(R,T)$ theories via Noether symmetry,''
Int.\ J.\ Geom.\ Meth.\ Mod.\ Phys.\  {\bf 12}, no. 10, 1550101 (2015),
%doi:10.1142/S0219887815501017
[arXiv:1502.00977 [gr-qc]].
%%CITATION = doi:10.1142/S0219887815501017;%%
%31 citations counted in INSPIRE as of 08 Jul 2017

%\cite{Myrzakulov:2015qaa}
%\bibitem{Myrzakulov:2015qaa} 
R.~Myrzakulov, L.~Sebastiani and S.~Vagnozzi,
%``Inflation in $f(R,\phi )$ -theories and mimetic gravity scenario,''
Eur.\ Phys.\ J.\ C {\bf 75}, 444 (2015),
%doi:10.1140/epjc/s10052-015-3672-6
[arXiv:1504.07984 [gr-qc]].
%%CITATION = doi:10.1140/epjc/s10052-015-3672-6;%%
%54 citations counted in INSPIRE as of 08 Jul 2017

%\cite{Haghani:2015zga}
%\bibitem{Haghani:2015zga} 
Z.~Haghani, S.~Shahidi and M.~Shiravand,
%``Energy conditions in mimetic-$f(R)$ gravity,''
arXiv:1507.07726 [gr-qc].
%%CITATION = ARXIV:1507.07726;%%
%14 citations counted in INSPIRE as of 08 Jul 2017

%\cite{Nojiri:2017ncd}
%\bibitem{Nojiri:2017ncd}
  S.~Nojiri, S.~D.~Odintsov and V.~K.~Oikonomou,
  %``Modified Gravity Theories on a Nutshell: Inflation, Bounce and Late-time Evolution,''
  arXiv:1705.11098 [gr-qc].
  %%CITATION = ARXIV:1705.11098;%%
  %15 citations counted in INSPIRE as of 07 Aug 2017

%\cite{Oikonomou:2016pkp}
%\bibitem{Oikonomou:2016pkp}
  V.~K.~Oikonomou,
  %``Aspects of Late-time Evolution in Mimetic $F(R)$ Gravity,''
  Mod.\ Phys.\ Lett.\ A {\bf 31} (2016) no.33,  1650191, 
%  doi:10.1142/S0217732316501911
  [arXiv:1609.03156 [gr-qc]].
  %%CITATION = doi:10.1142/S0217732316501911;%%
  %6 citations counted in INSPIRE as of 30 Sep 2017


%\cite{Nojiri:2016vhu}
%\bibitem{Nojiri:2016vhu}
  S.~Nojiri, S.~D.~Odintsov and V.~K.~Oikonomou,
  %``Viable Mimetic Completion of Unified Inflation-Dark Energy Evolution in Modified Gravity,''
  Phys.\ Rev.\ D {\bf 94} (2016) no.10,  104050
  doi:10.1103/PhysRevD.94.104050, 
  [arXiv:1608.07806 [gr-qc]].
  %%CITATION = doi:10.1103/PhysRevD.94.104050;%%
  %9 citations counted in INSPIRE as of 30 Sep 2017


%\cite{Odintsov:2016oyz}
%\bibitem{Odintsov:2016oyz}
  S.~D.~Odintsov and V.~K.~Oikonomou,
  %``Dark Energy Oscillations in Mimetic $F(R)$ Gravity,''
  Phys.\ Rev.\ D {\bf 94} (2016) no.4,  044012,
 % doi:10.1103/PhysRevD.94.044012, 
  [arXiv:1608.00165 [gr-qc]].
  %%CITATION = doi:10.1103/PhysRevD.94.044012;%%
  %8 citations counted in INSPIRE as of 30 Sep 2017

%\cite{Odintsov:2015wwp}
%\bibitem{Odintsov:2015wwp}
  S.~D.~Odintsov and V.~K.~Oikonomou,
  %``Accelerating cosmologies and the phase structure of F(R) gravity with Lagrange multiplier constraints: A mimetic approach,''
  Phys.\ Rev.\ D {\bf 93} (2016) no.2,  023517,
  %doi:10.1103/PhysRevD.93.023517, 
  [arXiv:1511.04559 [gr-qc]].
  %%CITATION = doi:10.1103/PhysRevD.93.023517;%%
  %26 citations counted in INSPIRE as of 30 Sep 2017

%\cite{Saadi:2014jfa}
%\bibitem{Saadi:2014jfa} 
  H.~Saadi,
  %``A Cosmological Solution to Mimetic Dark Matter,''
  Eur.\ Phys.\ J.\ C {\bf 76}, no. 1, 14 (2016), 
  %doi:10.1140/epjc/s10052-015-3856-0
  [arXiv:1411.4531 [gr-qc]].
  %%CITATION = doi:10.1140/epjc/s10052-015-3856-0;%%
  %13 citations counted in INSPIRE as of 18 Nov 2017

%\cite{Golovnev:2013jxa}
\bibitem{Golovnev:2013jxa} 
A.~Golovnev,
%``On the recently proposed Mimetic Dark Matter,''
Phys.\ Lett.\ B {\bf 728}, 39 (2014),
%doi:10.1016/j.physletb.2013.11.026
[arXiv:1310.2790 [gr-qc]].
%%CITATION = doi:10.1016/j.physletb.2013.11.026;%%
%55 citations counted in INSPIRE as of 01 Jul 2017

%\cite{Barvinsky:2013mea}
\bibitem{Barvinsky:2013mea} 
A.~O.~Barvinsky,
%``Dark matter as a ghost free conformal extension of Einstein theory,''
JCAP {\bf 1401}, 014 (2014),
%doi:10.1088/1475-7516/2014/01/014
[arXiv:1311.3111 [hep-th]].
%%CITATION = doi:10.1088/1475-7516/2014/01/014;%%
%38 citations counted in INSPIRE as of 01 Jul 2017

%\cite{Chaichian:2014qba}
%\bibitem{Chaichian:2014qba} 
M.~Chaichian, J.~Kluson, M.~Oksanen and A.~Tureanu,
%``Mimetic dark matter, ghost instability and a mimetic tensor-vector-scalar gravity,''
JHEP {\bf 1412}, 102 (2014),
%doi:10.1007/JHEP12(2014)102
[arXiv:1404.4008 [hep-th]].
%%CITATION = doi:10.1007/JHEP12(2014)102;%%
%34 citations counted in INSPIRE as of 08 Jul 2017

%\cite{Chamseddine:2014vna}
\bibitem{Chamseddine:2014vna} 
A.~H.~Chamseddine, V.~Mukhanov and A.~Vikman,
%``Cosmology with Mimetic Matter,''
JCAP {\bf 1406}, 017 (2014),
% doi:10.1088/1475-7516/2014/06/017
[arXiv:1403.3961 [astro-ph.CO]].
%%CITATION = doi:10.1088/1475-7516/2014/06/017;%%
%77 citations counted in INSPIRE as of 21 Jan 2017

%\cite{Mirzagholi:2014ifa}
\bibitem{Mirzagholi:2014ifa} 
L.~Mirzagholi and A.~Vikman,
%``Imperfect Dark Matter,''
JCAP {\bf 1506}, 028 (2015),
%doi:10.1088/1475-7516/2015/06/028
[arXiv:1412.7136 [gr-qc]].
%%CITATION = doi:10.1088/1475-7516/2015/06/028;%%
%31 citations counted in INSPIRE as of 28 May 2017

%\cite{Ijjas:2016pad}
\bibitem{Ijjas:2016pad} 
A.~Ijjas, J.~Ripley and P.~J.~Steinhardt,
%``NEC violation in mimetic cosmology revisited,''
Phys.\ Lett.\ B {\bf 760}, 132 (2016),
% doi:10.1016/j.physletb.2016.06.052
[arXiv:1604.08586 [gr-qc]].
%%CITATION = doi:10.1016/j.physletb.2016.06.052;%%
%3 citations counted in INSPIRE as of 21 Jan 2017

%\cite{Chamseddine:2016ktu}
\bibitem{Chamseddine:2016ktu} 
A.~H.~Chamseddine and V.~Mukhanov,
%``Nonsingular Black Hole,''
Eur.\ Phys.\ J.\ C {\bf 77}, no. 3, 183 (2017),
%doi:10.1140/epjc/s10052-017-4759-z
[arXiv:1612.05861 [gr-qc]].
%%CITATION = doi:10.1140/epjc/s10052-017-4759-z;%%
%6 citations counted in INSPIRE as of 23 Apr 2017

%\cite{Chamseddine:2016uef}
\bibitem{Chamseddine:2016uef} 
A.~H.~Chamseddine and V.~Mukhanov,
%``Resolving Cosmological Singularities,''
JCAP {\bf 1703}, no. 03, 009 (2017),
%doi:10.1088/1475-7516/2017/03/009
[arXiv:1612.05860 [gr-qc]].
%%CITATION = doi:10.1088/1475-7516/2017/03/009;%%
%10 citations counted in INSPIRE as of 23 Apr 2017

%\cite{Bodendorfer:2017bjt}
\bibitem{Bodendorfer:2017bjt} 
N.~Bodendorfer, A.~Schäfer and J.~Schliemann,
%``On the canonical structure of general relativity with a limiting curvature and its relation to loop quantum gravity,''
arXiv:1703.10670 [gr-qc].
%%CITATION = ARXIV:1703.10670;%%
%2 citations counted in INSPIRE as of 28 May 2017

%\cite{Liu:2017puc}
\bibitem{Liu:2017puc} 
D.~Langlois, K.~Noui, E.~Wilson-Ewing and H.~Liu,
%``Effective loop quantum cosmology as a higher-derivative scalar-tensor theory,''
Class.\ Quant.\ Grav.\  {\bf 34}, no. 22, 225004 (2017)
doi:10.1088/1361-6382/aa8f2f
[arXiv:1703.10812 [gr-qc]].
%%CITATION = doi:10.1088/1361-6382/aa8f2f;%%
%7 citations counted in INSPIRE as of 13 Nov 2017

\bibitem{LQC1} 
M.~Bojowald,
%``Loop quantum cosmology,''
Living Rev.\ Rel.\  {\bf 11}, 4 (2008).
%%CITATION = 00222,11,4;%%
%284 citations counted in INSPIRE as of 28 May 2017

%\cite{Ashtekar:2011ni}
\bibitem{LQC2} 
A.~Ashtekar and P.~Singh,
%``Loop Quantum Cosmology: A Status Report,''
Class.\ Quant.\ Grav.\  {\bf 28}, 213001 (2011),
%doi:10.1088/0264-9381/28/21/213001
[arXiv:1108.0893 [gr-qc]].
%%CITATION = doi:10.1088/0264-9381/28/21/213001;%%
%414 citations counted in INSPIRE as of 28 May 2017

%\cite{Banerjee:2011qu}
\bibitem{LQC3} 
K.~Banerjee, G.~Calcagni and M.~Martin-Benito,
%``Introduction to loop quantum cosmology,''
SIGMA {\bf 8}, 016 (2012),
%doi:10.3842/SIGMA.2012.016
[arXiv:1109.6801 [gr-qc]].
%%CITATION = doi:10.3842/SIGMA.2012.016;%%
%115 citations counted in INSPIRE as of 28 May 2017

%\cite{Mukhanov:1991zn}
\bibitem{Mukhanov:1991zn} 
V.~F.~Mukhanov and R.~H.~Brandenberger,
%``A Nonsingular universe,''
Phys.\ Rev.\ Lett.\  {\bf 68}, 1969 (1992).
%doi:10.1103/PhysRevLett.68.1969
%%CITATION = doi:10.1103/PhysRevLett.68.1969;%%
%173 citations counted in INSPIRE as of 20 Jun 2017

%\cite{Brandenberger:1993ef}
\bibitem{Brandenberger:1993ef} 
R.~H.~Brandenberger, V.~F.~Mukhanov and A.~Sornborger,
%``A Cosmological theory without singularities,''
Phys.\ Rev.\ D {\bf 48}, 1629 (1993),
%doi:10.1103/PhysRevD.48.1629
[gr-qc/9303001].
%%CITATION = doi:10.1103/PhysRevD.48.1629;%%
%155 citations counted in INSPIRE as of 20 Jun 2017

%\cite{Date:2008gq}
\bibitem{Date:2008gq} 
G.~Date and S.~Sengupta,
%``Effective Actions from Loop Quantum Cosmology: Correspondence with Higher Curvature Gravity,''
Class.\ Quant.\ Grav.\  {\bf 26}, 105002 (2009),
%doi:10.1088/0264-9381/26/10/105002
[arXiv:0811.4023 [gr-qc]].
%%CITATION = doi:10.1088/0264-9381/26/10/105002;%%
%9 citations counted in INSPIRE as of 20 Jun 2017

%\cite{Qiu:2011cy}
\bibitem{Qiu:2011cy} 
T.~Qiu, J.~Evslin, Y.~F.~Cai, M.~Li and X.~Zhang,
%``Bouncing Galileon Cosmologies,''
JCAP {\bf 1110}, 036 (2011),
%doi:10.1088/1475-7516/2011/10/036
[arXiv:1108.0593 [hep-th]].
%%CITATION = doi:10.1088/1475-7516/2011/10/036;%%
%116 citations counted in INSPIRE as of 20 Jun 2017

%\cite{Yoshida:2017swb}
\bibitem{Yoshida:2017swb} 
D.~Yoshida, J.~Quintin, M.~Yamaguchi and R.~H.~Brandenberger,
%``Cosmological perturbations and stability of nonsingular cosmologies with limiting curvature,''
Phys.\ Rev.\ D {\bf 96}, no. 4, 043502 (2017),
%doi:10.1103/PhysRevD.96.043502
[arXiv:1704.04184 [hep-th]].
%%CITATION = doi:10.1103/PhysRevD.96.043502;%%
%5 citations counted in INSPIRE as of 13 Nov 2017

%\cite{Firouzjahi:2017txv}
\bibitem{Firouzjahi:2017txv} 
H.~Firouzjahi, M.~A.~Gorji and S.~A.~Hosseini Mansoori,
%``Instabilities in Mimetic Matter Perturbations,''
JCAP {\bf 1707}, 031 (2017),
%doi:10.1088/1475-7516/2017/07/031
[arXiv:1703.02923 [hep-th]].
%%CITATION = doi:10.1088/1475-7516/2017/07/031;%%
%11 citations counted in INSPIRE as of 13 Nov 2017

%\cite{Hirano:2017zox}
\bibitem{Hirano:2017zox} 
S.~Hirano, S.~Nishi and T.~Kobayashi,
%``Healthy imperfect dark matter from effective theory of mimetic cosmological perturbations,''
JCAP {\bf 1707}, no. 07, 009 (2017)
doi:10.1088/1475-7516/2017/07/009
[arXiv:1704.06031 [gr-qc]].
%%CITATION = doi:10.1088/1475-7516/2017/07/009;%%
%9 citations counted in INSPIRE as of 13 Nov 2017

%\cite{Cai:2017dyi}
\bibitem{Cai:2017dyi} 
Y.~Cai and Y.~S.~Piao,
%``A covariant Lagrangian for stable nonsingular bounce,''
JHEP {\bf 1709}, 027 (2017)
doi:10.1007/JHEP09(2017)027
[arXiv:1705.03401 [gr-qc]].
%%CITATION = doi:10.1007/JHEP09(2017)027;%%
%10 citations counted in INSPIRE as of 13 Nov 2017

%\cite{Cai:2017dxl}
\bibitem{Cai:2017dxl} 
Y.~Cai and Y.~S.~Piao,
%``Higher order derivative coupling to gravity and its cosmological implications,''
arXiv:1707.01017 [gr-qc].
%%CITATION = ARXIV:1707.01017;%%

%\cite{Zheng:2017qfs}
\bibitem{Zheng:2017qfs} 
Y.~Zheng, L.~Shen, Y.~Mou and M.~Li,
%``On (in)stabilities of perturbations in mimetic models with higher derivatives,''
JCAP {\bf 1708}, no. 08, 040 (2017)
doi:10.1088/1475-7516/2017/08/040
[arXiv:1704.06834 [gr-qc]].
%%CITATION = doi:10.1088/1475-7516/2017/08/040;%%
%7 citations counted in INSPIRE as of 13 Nov 2017

%\cite{Horndeski:1974wa}
\bibitem{Horndeski:1974wa} 
G.~W.~Horndeski,
%``Second-order scalar-tensor field equations in a four-dimensional space,''
Int.\ J.\ Theor.\ Phys.\  {\bf 10}, 363 (1974).
%doi:10.1007/BF01807638
%%CITATION = doi:10.1007/BF01807638;%%
%678 citations counted in INSPIRE as of 02 Jul 2017

%\cite{Nicolis:2008in}
\bibitem{Nicolis:2008in} 
A.~Nicolis, R.~Rattazzi and E.~Trincherini,
%``The Galileon as a local modification of gravity,''
Phys.\ Rev.\ D {\bf 79}, 064036 (2009),
%doi:10.1103/PhysRevD.79.064036
[arXiv:0811.2197 [hep-th]].
%%CITATION = doi:10.1103/PhysRevD.79.064036;%%
%985 citations counted in INSPIRE as of 02 Jul 2017

%\cite{Deffayet:2009wt}
\bibitem{Deffayet:2009wt} 
C.~Deffayet, G.~Esposito-Farese and A.~Vikman,
%``Covariant Galileon,''
Phys.\ Rev.\ D {\bf 79}, 084003 (2009),
%doi:10.1103/PhysRevD.79.084003
[arXiv:0901.1314 [hep-th]].
%%CITATION = doi:10.1103/PhysRevD.79.084003;%%
%555 citations counted in INSPIRE as of 02 Jul 2017

%\cite{Ostrogradsky:1850fid}
\bibitem{Ostrogradsky:1850fid} 
M.~Ostrogradsky,
%``Mémoires sur les équations différentielles, relatives au problème des isopérimètres,''
Mem.\ Acad.\ St.\ Petersbourg {\bf 6}, no. 4, 385 (1850).
%18 citations counted in INSPIRE as of 02 Jul 2017

%\cite{Woodard:2015zca}
\bibitem{Woodard:2015zca} 
R.~P.~Woodard,
%``Ostrogradsky's theorem on Hamiltonian instability,''
Scholarpedia {\bf 10}, no. 8, 32243 (2015),
%doi:10.4249/scholarpedia.32243
[arXiv:1506.02210 [hep-th]].
%%CITATION = doi:10.4249/scholarpedia.32243;%%
%91 citations counted in INSPIRE as of 02 Jul 2017

%\cite{Gleyzes:2014dya}
\bibitem{Gleyzes:2014dya} 
J.~Gleyzes, D.~Langlois, F.~Piazza and F.~Vernizzi,
%``Healthy theories beyond Horndeski,''
Phys.\ Rev.\ Lett.\  {\bf 114}, no. 21, 211101 (2015),
%doi:10.1103/PhysRevLett.114.211101
[arXiv:1404.6495 [hep-th]].
%%CITATION = doi:10.1103/PhysRevLett.114.211101;%%
%195 citations counted in INSPIRE as of 02 Jul 2017

%\cite{Langlois:2015cwa}
\bibitem{Langlois:2015cwa} 
D.~Langlois and K.~Noui,
%``Degenerate higher derivative theories beyond Horndeski: evading the Ostrogradski instability,''
JCAP {\bf 1602}, no. 02, 034 (2016),
%doi:10.1088/1475-7516/2016/02/034
[arXiv:1510.06930 [gr-qc]].
%%CITATION = doi:10.1088/1475-7516/2016/02/034;%%
%70 citations counted in INSPIRE as of 02 Jul 2017

%\cite{Crisostomi:2016czh}
\bibitem{Crisostomi:2016czh} 
  M.~Crisostomi, K.~Koyama and G.~Tasinato,
  %``Extended Scalar-Tensor Theories of Gravity,''
  JCAP {\bf 1604}, no. 04, 044 (2016), 
 % doi:10.1088/1475-7516/2016/04/044
  [arXiv:1602.03119 [hep-th]].
  %%CITATION = doi:10.1088/1475-7516/2016/04/044;%%
  %45 citations counted in INSPIRE as of 18 Nov 2017
  
%\cite{Crisostomi:2016tcp}
\bibitem{Crisostomi:2016tcp} 
  M.~Crisostomi, M.~Hull, K.~Koyama and G.~Tasinato,
  %``Horndeski: beyond, or not beyond?,''
  JCAP {\bf 1603}, no. 03, 038 (2016)
 % doi:10.1088/1475-7516/2016/03/038
  [arXiv:1601.04658 [hep-th]].
  %%CITATION = doi:10.1088/1475-7516/2016/03/038;%%
  %44 citations counted in INSPIRE as of 18 Nov 2017





%\cite{Arroja:2015yvd}
\bibitem{Arroja:2015yvd} 
F.~Arroja, N.~Bartolo, P.~Karmakar and S.~Matarrese,
%``Cosmological perturbations in mimetic Horndeski gravity,''
JCAP {\bf 1604}, no. 04, 042 (2016),
%doi:10.1088/1475-7516/2016/04/042
[arXiv:1512.09374 [gr-qc]].
%%CITATION = doi:10.1088/1475-7516/2016/04/042;%%
%12 citations counted in INSPIRE as of 08 Jul 2017

%\cite{Cognola:2016gjy}
\bibitem{Cognola:2016gjy} 
G.~Cognola, R.~Myrzakulov, L.~Sebastiani, S.~Vagnozzi and S.~Zerbini,
%``Covariant Hořava-like and mimetic Horndeski gravity: cosmological solutions and perturbations,''
Class.\ Quant.\ Grav.\  {\bf 33}, no. 22, 225014 (2016),
%doi:10.1088/0264-9381/33/22/225014
[arXiv:1601.00102 [gr-qc]].
%%CITATION = doi:10.1088/0264-9381/33/22/225014;%%
%21 citations counted in INSPIRE as of 08 Jul 2017

%\cite{Arroja:2017msd}
\bibitem{Arroja:2017msd} 
F.~Arroja, T.~Okumura, N.~Bartolo, P.~Karmakar and S.~Matarrese,
``Large-scale structure in mimetic Horndeski gravity,''
arXiv:1708.01850 [astro-ph.CO].
%%CITATION = ARXIV:1708.01850;%%
%2 citations counted in INSPIRE as of 02 Sep 2017

%\cite{Takahashi:2017pje}
\bibitem{Takahashi:2017pje} 
K.~Takahashi and T.~Kobayashi,
``Extended mimetic gravity: Hamiltonian analysis and gradient instabilities,''
arXiv:1708.02951 [gr-qc].
%%CITATION = ARXIV:1708.02951;%%
%1 citations counted in INSPIRE as of 02 Sep 2017

%\cite{BenAchour:2016fzp}
\bibitem{BenAchour:2016fzp} 
J.~Ben Achour, M.~Crisostomi, K.~Koyama, D.~Langlois, K.~Noui and G.~Tasinato,
%``Degenerate higher order scalar-tensor theories beyond Horndeski up to cubic order,''
JHEP {\bf 1612}, 100 (2016),
%doi:10.1007/JHEP12(2016)100
[arXiv:1608.08135 [hep-th]].
%%CITATION = doi:10.1007/JHEP12(2016)100;%%
%23 citations counted in INSPIRE as of 22 Jun 2017

%\cite{Langlois:2015skt}
\bibitem{Langlois:2015skt} 
D.~Langlois and K.~Noui,
%``Hamiltonian analysis of higher derivative scalar-tensor theories,''
JCAP {\bf 1607}, no. 07, 016 (2016),
%doi:10.1088/1475-7516/2016/07/016
[arXiv:1512.06820 [gr-qc]].
%%CITATION = doi:10.1088/1475-7516/2016/07/016;%%
%42 citations counted in INSPIRE as of 22 Jun 2017

%\cite{Langlois:2017mxy}
\bibitem{Langlois:2017mxy} 
D.~Langlois, M.~Mancarella, K.~Noui and F.~Vernizzi,
%``Effective Description of Higher-Order Scalar-Tensor Theories,''
JCAP {\bf 1705}, no. 05, 033 (2017), 
%doi:10.1088/1475-7516/2017/05/033
[arXiv:1703.03797 [hep-th]].
%%CITATION = doi:10.1088/1475-7516/2017/05/033;%%
%6 citations counted in INSPIRE as of 22 Jun 2017

%\cite{Nojiri:2014zqa}
\bibitem{Nojiri:2014zqa} 
S.~Nojiri and S.~D.~Odintsov,
%``Mimetic $F(R)$ gravity: inflation, dark energy and bounce,''
Mod.\ Phys.\ Lett.\ A {\bf 29}, no. 40, 1450211 (2014),
%doi:10.1142/S0217732314502113
[arXiv:1408.3561 [hep-th]].
%%CITATION = doi:10.1142/S0217732314502113;%%
%53 citations counted in INSPIRE as of 08 Jul 2017

%\cite{DeFelice:2015moy}
\bibitem{DeFelice:2015moy} 
  A.~De Felice and S.~Mukohyama,
  %``Phenomenology in minimal theory of massive gravity,''
  JCAP {\bf 1604}, no. 04, 028 (2016),
  %doi:10.1088/1475-7516/2016/04/028
  [arXiv:1512.04008 [hep-th]].
  %%CITATION = doi:10.1088/1475-7516/2016/04/028;%%
  %11 citations counted in INSPIRE as of 27 Feb 2017

%\cite{Gumrukcuoglu:2016jbh}
\bibitem{Gumrukcuoglu:2016jbh} 
  A.~E.~ Gumrukcuoglu, S.~Mukohyama and T.~P.~Sotiriou,
  %``Low energy ghosts and the Jeans? instability,''
  Phys.\ Rev.\ D {\bf 94}, no. 6, 064001 (2016), 
  %doi:10.1103/PhysRevD.94.064001
  [arXiv:1606.00618 [hep-th]].
  %%CITATION = doi:10.1103/PhysRevD.94.064001;%%
  %5 citations counted in INSPIRE as of 27 Feb 2017

%\cite{Babichev:2016jzg}
\bibitem{Babichev:2016jzg} 
  E.~Babichev and S.~Ramazanov,
  %``Gravitational focusing of Imperfect Dark Matter,''
  Phys.\ Rev.\ D {\bf 95}, no. 2, 024025 (2017), 
  %doi:10.1103/PhysRevD.95.024025
  [arXiv:1609.08580 [gr-qc]].
  %%CITATION = doi:10.1103/PhysRevD.95.024025;%%
  %3 citations counted in INSPIRE as of 06 May 2017

%\cite{Babichev:2017lrx}
\bibitem{Babichev:2017lrx} 
E.~Babichev and S.~Ramazanov,
%``Caustic free completion of pressureless perfect fluid and k-essence,''
JHEP {\bf 1708}, 040 (2017)
doi:10.1007/JHEP08(2017)040
[arXiv:1704.03367 [hep-th]].
%%CITATION = doi:10.1007/JHEP08(2017)040;%%
%7 citations counted in INSPIRE as of 13 Nov 2017

%\cite{Randall:1999ee}
\bibitem{Randall:1999ee} 
  L.~Randall and R.~Sundrum,
  %``A Large mass hierarchy from a small extra dimension,''
  Phys.\ Rev.\ Lett.\  {\bf 83}, 3370 (1999), 
  %doi:10.1103/PhysRevLett.83.3370
  [hep-ph/9905221].\\
  %%CITATION = doi:10.1103/PhysRevLett.83.3370;%%
  %7651 citations counted in INSPIRE as of 14 Sep 2017
%\cite{Randall:1999vf}
%\bibitem{Randall:1999vf} 
  L.~Randall and R.~Sundrum,
  %``An Alternative to compactification,''
  Phys.\ Rev.\ Lett.\  {\bf 83}, 4690 (1999), 
  %doi:10.1103/PhysRevLett.83.4690
  [hep-th/9906064].
  %%CITATION = doi:10.1103/PhysRevLett.83.4690;%%
  %6009 citations counted in INSPIRE as of 14 Sep 2017

%\cite{Cline:1999ts}
\bibitem{Cline:1999ts} 
  J.~M.~Cline, C.~Grojean and G.~Servant,
  %``Cosmological expansion in the presence of extra dimensions,''
  Phys.\ Rev.\ Lett.\  {\bf 83}, 4245 (1999)
 % doi:10.1103/PhysRevLett.83.4245
  [hep-ph/9906523].
  %%CITATION = doi:10.1103/PhysRevLett.83.4245;%%
  %608 citations counted in INSPIRE as of 14 Sep 2017


%\cite{Maartens:2010ar}
\bibitem{Maartens:2010ar} 
R.~Maartens and K.~Koyama,
%``Brane-World Gravity,''
Living Rev.\ Rel.\  {\bf 13}, 5 (2010)
doi:10.12942/lrr-2010-5
[arXiv:1004.3962 [hep-th]].
%%CITATION = doi:10.12942/lrr-2010-5;%%
%283 citations counted in INSPIRE as of 21 Jun 2017

%\cite{Brizuela:2008ra}
\bibitem{Brizuela:2008ra} 
D.~Brizuela, J.~M.~Martin-Garcia and G.~A.~Mena Marugan,
%``xPert: Computer algebra for metric perturbation theory,''
Gen.\ Rel.\ Grav.\  {\bf 41}, 2415 (2009), 
%doi:10.1007/s10714-009-0773-2
[arXiv:0807.0824 [gr-qc]].
%%CITATION = doi:10.1007/s10714-009-0773-2;%%
%38 citations counted in INSPIRE as of 07 Jul 2017

%\cite{Nutma:2013zea}
\bibitem{Nutma:2013zea} 
T.~Nutma,
%``xTras : A field-theory inspired xAct  package for mathematica,''
Comput.\ Phys.\ Commun.\  {\bf 185}, 1719 (2014), 
%doi:10.1016/j.cpc.2014.02.006
[arXiv:1308.3493 [cs.SC]].
%%CITATION = doi:10.1016/j.cpc.2014.02.006;%%
%38 citations counted in INSPIRE as of 07 Jul 2017


\end{thebibliography}
\end{document}